\documentclass[prd,nofootinbib,showpacs,preprint]{revtex4}
\usepackage{latexsym}
\usepackage{dcolumn}
\usepackage{epsfig}

\begin{document}

\preprint{FERMILAB-Pub-04-142-T}

\title{Understanding single-top-quark production and jets\\
at hadron colliders}

\author{Zack~Sullivan}
\affiliation{Theoretical Physics Department, Fermi National Accelerator
Laboratory,\\ Batavia, IL 60510-0500, USA}

\date{August 3, 2004}

\begin{abstract}
I present an analysis of fully differential single-top-quark
production plus jets at next-to-leading order.  I describe the effects
of jet definitions, top-quark mass, and higher orders on the shapes
and normalizations of the kinematic distributions, and quantify all
theoretical uncertainties.  I explain how to interpret
next-to-leading-order jet calculations, and compare them to showering
event generators.  Using the program ZTOP, I show that HERWIG and
PYTHIA significantly underestimate both $s$-channel and $t$-channel
single-top-quark production, and propose a scheme to match the
relevant samples to the next-to-leading-order predictions.
\end{abstract}


\maketitle

\section{Introduction}
\label{sec:introduction}

The discovery of single-top-quark production will be the flagship
measurement of run II at the Fermilab Tevatron.  Preliminary run II
results \cite{Juste:2004an} are already surpassing the limits set by
the CDF \cite{CDF} and D0 \cite{D0} Collaborations using data from run
I.  Over the next few years, the single-top-quark cross sections will
be accurately measured at both the Fermilab Tevatron \cite{tev2000}
and the CERN Large Hadron Collider (LHC) \cite{Beneke:2000hk}.  In
order to compare the coming measurements of single-top-quark
production to theory, a detailed understanding of the predictions for
jet distributions, and the associated uncertainties is required.

\begin{figure}[tbh]
\epsfxsize=3.125in 
\centerline{\epsfbox{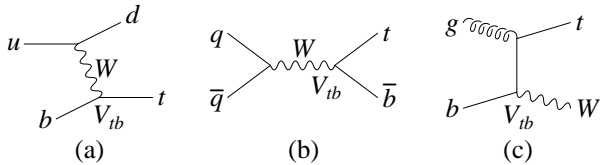}}
\caption{Representative leading-order Feynman diagrams for (a) $t$-channel,
(b) $s$-channel, and (c) $Wt$-associated production of a single top
quark.  The CKM matrix element $V_{tb}$ appears directly in the
production diagrams.
\label{fig:prodvtb}}
\end{figure}

The measurement of single-top-quark production will provide an
excellent opportunity to study the charged-current weak-interaction of
the top quark.  Within the standard model, a measurement of the cross
section allows direct extraction of the Cabibbo-Kobayashi-Maskawa
(CKM) matrix element $|V_{tb}|^2$.  In Fig.~\ref{fig:prodvtb} we see
that $V_{tb}$ appears in the leading-order (LO) Feynman diagrams for
$t$-channel production
\cite{Willenbrock:cr,Yuan:1989tc,Ellis:1992yw,Carlson:1993dt,Bordes:1995ki,Heinson:1996zm,Stelzer:1997ns,Harris:2002md},
$s$-channel production
\cite{Cortese:fw,Stelzer:1995mi,Smith:1996ij,Mrenna:1997wp,Harris:2000sv,Harris:2002md},
and $Wt$-associated production
\cite{Heinson:1996zm,Moretti:1997ng,Tait:1999cf,Belyaev:2000me,Zhu:2001hw}.
This paper focuses on $s$-channel and $t$-channel production, which
have sizable cross sections, and may be distinguished experimentally
by the number of $b$ tags in the final state \cite{Stelzer:1998ni}.
Since the $Wt$-associated production cross section is small at the
Tevatron, and no differential next-to-leading-order calculation
exists, $Wt$ production is not addressed here.

Models that extend the standard model often predict large corrections
to $s$-channel or $t$-channel production, but not to both
\cite{Tait:2000sh}.  Anything that produces an anomalous coupling
\cite{Carlson:1994bg,Kane:1991bg,Rizzo:1995uv,Tait:1996dv,Datta:1996gg,Whisnant:1997qu,Boos:1999dd,Tait:2000sh,Espriu:2001vj},
or flavor changing neutral current \cite{Tait:2000sh,Han:1998tp},
between the top quark and any other quark opens new production modes
in the $t$-channel, but only slightly reduces the fraction of $b$ jets
in $s$-channel production.  Conversely, any process that allows a new
intermediate particle, such as strong dynamics
\cite{Simmons:1996ws,Lu:1997aa,Baringer:1997wu,Yue:ff}, a charged
top-pion \cite{Yue:1997nf,He:1998ie,Tait:2000sh,Cao:2002af},
Kaluza-Klein modes of the $W$ \cite{Datta:2000gm}, or a $W^\prime$
boson \cite{Simmons:1996ws,Li:nk,Malkawi:1996fs,Muller:1996dj,Tait:2000sh},
would enhance $s$-channel production, but is highly mass suppressed in
$t$-channel production.  In fact, $s$-channel single-top-quark
production has been found to be the most sensitive probe of high-mass
charged vector $W^{\prime}$ bosons \cite{Sullivan:2002jt} at the
Tevatron \cite{Acosta:2002nu} and LHC \cite{Sullivan:2003xy}.

Both $s$-channel and $t$-channel production are important in direct
searches for CP violation
\cite{Atwood:1996pd,Bar-Shalom:1997si,Christova:2001zm},
$R$-parity-conserving supersymmetry \cite{Li:1996bh}, and
$R$-parity-violating supersymmetry
\cite{Datta:1997us,Oakes:1997zg,Hikasa:1999wy,Chiappetta:1999cd,Allanach:1999bf}.
Both channels also contribute a significant background to all signals
that include $W+\mathrm{jets}$ with, or without, $b$ tags.  These
backgrounds appear in a number of Higgs search channels
\cite{Carena:2000yx} and other new physics, such as
$R$-parity-conserving \cite{Abel:2000vs} and $R$-parity-violating
\cite{Berger:1999zt} supersymmetry searches.

The analytic form of the fully differential next-to-leading-order
(NLO) cross section for $s$-channel and $t$-channel production of a
single top-quark plus one jet appears in Ref.~\cite{Harris:2002md}.
The spin-averaged version of that calculation is available from this
author in the form of a FORTRAN program called ZTOP, and is used in
this paper.  A recurring theme throughout the analysis of the
kinematic distributions is that the next-to-leading-order calculation
describes \emph{jets}, and not partons.  This affects the questions
that may be asked about the final state, and contributes additional
challenges to the matching between theory and experiment.

The ultimate goal of performing differential calculations at
next-to-leading order is to improve the quantitative connection
between theory and experiment.  As important as the distributions
themselves, is a reliable estimate of the uncertainties in these
distributions.  Previous estimates of the uncertainties in the parton
luminosities and kinematics \cite{Stelzer:1998ni} have been somewhat
rough, and are revisited with more careful analysis.  Analytic
calculations are not typically compared to raw data, but rather to
event generators, such as HERWIG \cite{Corcella:2000bw} and PYTHIA
\cite{Sjostrand:2000wi}, that have been tuned to fit the data.  It is
vital to determine whether these event generators at least predict the
shapes of the measured distributions accurately.  We will see that the
current event generators provide very poor predictions of both
$s$-channel and $t$-channel production.

This paper is organized as follows.  A general search strategy for
discovering single-top-quark production appears in
Sec.~\ref{sec:discovery}.  This lays the groundwork for discussing the
event rates in terms of the number of jets in the final state.  The
next-to-leading-order distributions for both $s$-channel and
$t$-channel single-top-quark production at the Tevatron are presented
in Sec.~\ref{sec:nlopred}.  The hazards of treating NLO calculations as
parton calculations are explored, and subtleties in the interpretation
of jet calculations are discussed in detail.  The bottom-quark mass is
shown to have no measurable effect on the distributions.  All
uncertainties in the inclusive and exclusive theoretical calculations
are described in Sec.~\ref{sec:uncertainties}.  In
Sec.~\ref{sec:eventgen} the PYTHIA \cite{Sjostrand:2000wi} and HERWIG
\cite{Corcella:2000bw} event generators are shown to have poor
representations of the measurable kinematic distributions.  In
Sec.~\ref{sec:gencor} a simple scheme is proposed to match the event
generators to the next-to-leading-order calculation.  The paper
concludes in Sec.~\ref{sec:conclusions} with a few directions for
improvement.

\subsection{Strategy for discovery}
\label{sec:discovery}

The final state for both $s$- and $t$-channel production of single top
quarks contains two jets plus the decay of a $W$ boson.  In order to
extract the signal from the backgrounds, the decay of the $W$ to an
electron or muon, and a neutrino is used.  The most serious background
for single-top-quark production is the production of $t\bar t$ pairs
\cite{Stelzer:1998ni}.  The number of $t\bar t$ pairs that will be
reconstructed in detectors at the Tevatron is about 5 times the total
number of single-top-quark events in the 1 $b$-tag sample, and about
10 times the number in the 2 $b$-tag sample.  Because there is already
a top quark in the final state, the essence of any study will be to
reduce the $t\bar t$ background, while maintaining a measurable
signal.

The most effective method to discover single-top-quark production uses
a ``jet veto,'' which was first proposed in
Ref.~\cite{Stelzer:1998ni}.  This method relies on the observation
that in $t\bar t$ production, the extra top quark tends to produce
additional high-$E_T$ objects (leptons or jets) in the final state.
The basic strategy is to form a sample of two jets, a lepton, and
missing transverse energy $E_T$.  If there are any additional isolated
``hard'' jets or leptons then the event is rejected.  The term
``hard'' can be loosely defined to refer to any jet or lepton with
$E_T > 15$--$25$ GeV, and $|\eta| < 2.5$--$4$.  Experimentally, the
``jet veto'' is identical to separating the final states into a $W+2$
jet exclusive sample and a $W+3$ jet inclusive sample.

We observe in Fig.\ 2 of Ref.~\cite{Stelzer:1998ni} that placing cuts
anywhere in the region of $E_{Tj}=15$--$25$ GeV for every jet yields
the same significance.  The reason is that acceptance for events that
contain an extra radiated jet in the signal increases as the cut is
lowered (thereby failing to pass the cuts) at the same rate that the
acceptance for the highest-$E_T$ jet increases.  Hence, in order to
avoid a sharp drop in efficiency, the jet veto should be set at the
same $E_T$ as the cut on the second-highest-$E_T$ jet.

The largest remaining background is $Wjj$ production --- dominantly
$Wcj+Wjj$ in the 1 $b$-tag sample, and $Wb\bar b$ in the 2 $b$-tag
sample.  This background decreases as a function of reconstructed
top-quark mass.  Hence, a top-quark mass cut reduces this to an
acceptable level.  If a top-quark mass cut cannot be used, e.g.,
because of poor jet-energy resolution, other discriminates will have
to be found.  However, these will be ineffective without the
corrections described in Sec.~\ref{sec:eventgen}.

The detection strategies of Refs.~\cite{tev2000,Stelzer:1998ni} should
be merged into a program of three successive measurements.  The first
stage measures the $W+2$ jet signal with at least 1 $b$ tag.  This
sums over all single-top-quark production modes, and requires the
least amount of integrated luminosity for discovery.  As data is
accumulated, the samples can be split into independent subsets with 1
or 2 $b$ tags.  The 1 $b$-tag sample can be used to study $t$-channel
production, and the 2 $b$-tag sample can be used to study $s$-channel
production.

\section{Next-to-leading-order distributions}
\label{sec:nlopred}

Fully differential next-to-leading-order cross sections predict
\emph{jet} distributions; they do not predict parton distributions.
The consequences of this can be subtle, and require care when
comparing fixed-order calculations with data, or with showering event
generators.  To understand these comparisons, only distributions for
top-quark production at the Tevatron are examined.  Distributions for
antitop-quark production at the Tevatron differ solely by the signs of
the pseudorapidities, and figures would be redundant.  Since the
physics issues are identical, no jet distributions for the LHC are
shown.

The challenge of calculating differential cross sections is in how to
deal with final-state infrared singularities.  Different methods have
different solutions in detail, but they all impose one of two
conditions: include a jet definition, or include a hadronization
function.  The first consequence of including a jet definition in the
base theoretical calculation is that the same choice of jet definition
must be used in any comparison to data.  This may appear obvious, but
it contains a subtlety: how do you compare a jet that consists of a
few partons with one that contains multiple particles?  In
Sec.~\ref{sec:jetdef} we will see the effect of the choice of jet
definition on theoretical distributions.  Comparisons between theory
and experiment are actually made at the event generator level, rather
than directly with data.  Hence, in Sec.~\ref{sec:eventgen} we will
adopt an effective solution for matching the NLO calculations to the
event generators.

\begin{figure}[htb]
\epsfxsize=3.125in 
\centerline{\epsfbox{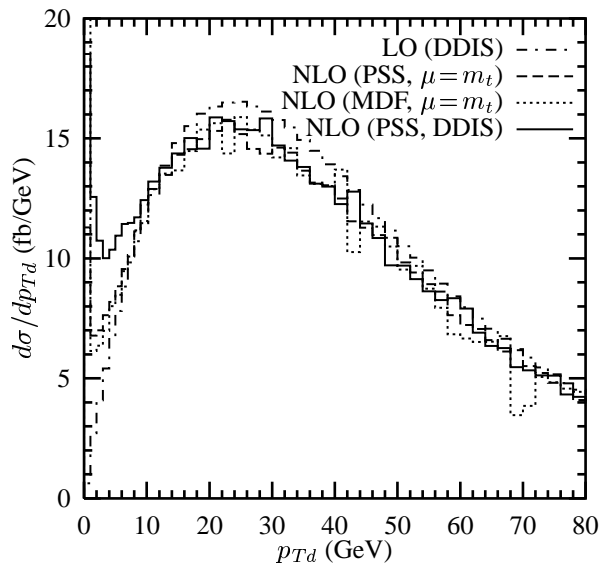}}
\caption{Transverse momentum $p_{Td}$ of the ``$d$''-jet in $t$-channel
production at leading and next-to-leading order for the phase space
slicing (PSS) and massive dipole formalism (MDF) calculations, and two
scales, $m_t$ and the double deep-inelastic-scattering scales ($Q^2$
and $Q^2+m_t^2$).  Leading order includes a $K$-factor of $1.09$.
\label{fig:nlod}}
\end{figure}

Before presenting NLO distributions, we must first understand which
distributions are appropriate to calculate.  We might be tempted to
look at the LO diagrams and ask for the corresponding NLO
distributions, but this would not be useful in general.  Consider the
LO diagram for $t$-channel production in Fig.~\ref{fig:prodvtb}(a).
If we plot the NLO transverse momentum of the jet that includes the
final-state $d$ quark, we see in Fig.~\ref{fig:nlod} that the
calculation is not very stable.  The massive dipole formalism (MDF)
and phase space slicing (PSS) calculations of
Ref.~\cite{Harris:2002md} provide the same results at the same scale.
The methods only differ in the bin that exactly includes $p_{Td} = 0$
GeV, which is an artifact of the different subtractions methods, and
in any case not observable.

The problem with Fig.~\ref{fig:nlod} is that the shapes of the
distributions are not stable compared to scale variation or
perturbative order.  At leading order the $d$ quark recoils against
the massive top quark, and so has a finite distribution at small
$p_{Td}$.  At NLO the distribution takes an arbitrary value that
depends on the choice of scale.  Also the shapes at larger
(i.e. measurable) $p_{Td}$ change between orders, which leads to the
false impression that the NLO jets are softer than the LO parton.
This figure demonstrates the danger of interpreting \emph{jet} results
in \emph{partonic} language.

\begin{figure}[htb]
\epsfxsize=3.125in 
\centerline{\epsfbox{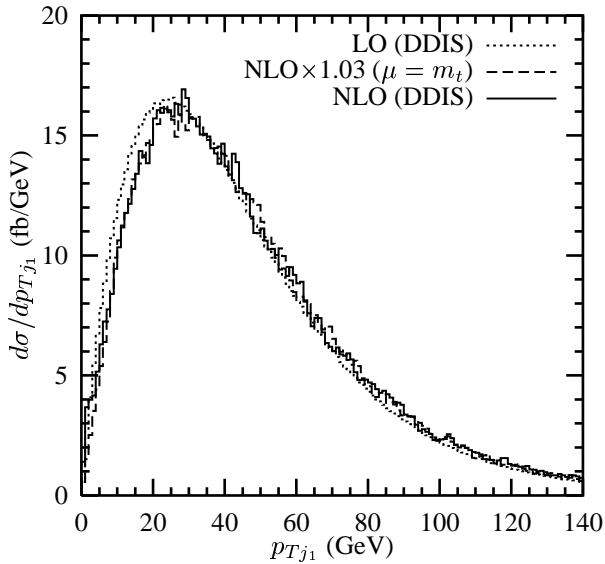}}
\caption{Transverse momentum $p_{Tj_1}$ of the highest-$p_T$ jet in
$t$-channel production at leading and next-to-leading order for
different scales.
\label{fig:nloj1}}
\end{figure}

The distribution that should have been plotted is the transverse
momentum of the highest-$p_T$ jet.  In Fig.~\ref{fig:nloj1} we see
that this distribution is stable in shape at different orders in
perturbation theory, and for different scales.  In fact the difference
between the LO and NLO distribution is smaller than the NLO scale
uncertainty.  We also see that the choice of scale effects only
normalization.  Hence, we have a very well-predicted quantity.

What stabilizes the $p_T$-ordered jet at NLO if the ``$d$'' jet is so
unstable?  The answer is that when the $d$ quark becomes softer, the
parton that recoils against the top quark is a $\bar u$ that came from
initial-state splitting of a gluon.  Hence, the kinematically stable
jet has a different flavor content than the LO jet.  We are fortunate
that we cannot distinguish experimentally the difference between a jet
that started from an up or down quark.  If the jet is produced with a
charm or bottom quark, however, we have to track it explicitly because
it might be taggable.  In the case of $t$-channel single-top-quark
production, this is CKM-suppressed and occurs less than 1\% of the
time in the light-quark initiated side of the diagram.  Once we look
at $p_T$-ordered jets, we will have to keep track of how often the
leading jet includes a $b$ quark from the other incoming hadron.  This
will not cause any stability problems, however, since there are no
overlapping singularities between the light-quark and heavy-quark
initiated parts of the diagrams.

\subsection{$t$-channel NLO distributions}
\label{sec:etvsnlo}

\begin{figure}[tbh]
\epsfxsize=3.125in 
\centerline{\epsfbox{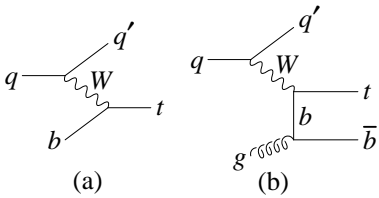}}
\caption{Representative leading-order Feynman diagrams for (a) $t$-channel
production ($qb\to tq^\prime$), and (b) $W$-gluon fusion ($qg\to t\bar
bq^\prime$).
\label{fig:prodt}}
\end{figure}

We have seen the stability of the distribution of the highest-$p_T$
jet between leading and next-to-leading order.  There are two
historically relevant questions.  First, how well are the leading jet
and top quark distributions modeled by $qb\to tq^\prime$
[Fig.~\ref{fig:prodt}(a)] or $qg\to t\bar bq^\prime$
[Fig.~\ref{fig:prodt}(b)] parton-level calculations?  Second, how do
typical cuts effect the distributions of the top quark and jets?  The
analysis of leading-order $qb\to tq^\prime$ production plus showering
is postponed until Sec.~\ref{sec:eventgen}, when we observe the
failure of the current event generators to reproduce the jet spectrum.

The analytic form of the $t$-channel cross section has the same form
as double deep-inelastic-scattering (DDIS)
\cite{Stelzer:1997ns,Harris:2002md}.  Color conservation enforces a
natural factorization of the scales.  The fermion line in
Fig.~\ref{fig:prodt}(a) that does not include a top quark probes a
proton with the DIS scale $Q^2$, which is identical to the virtuality
of the $W$ boson through NLO.  The fermion line that does connect to a
top quark sees the DIS scale for massive quarks of $Q^2+m_t^2$.
Another reasonable choice for both scales is the mass of the top
quark.  The difference between using the DDIS scales, or the top-quark
mass is less than 3\% at next-to-leading order.  But the cross section
and distributions do not change from LO to NLO when the DDIS scales
are used.  This is an indication of the relation of this calculation
to the extraction of parton distribution functions, and DIS in general.

\begin{figure}[tbh]
\epsfxsize=3.125in 
\centerline{\epsfbox{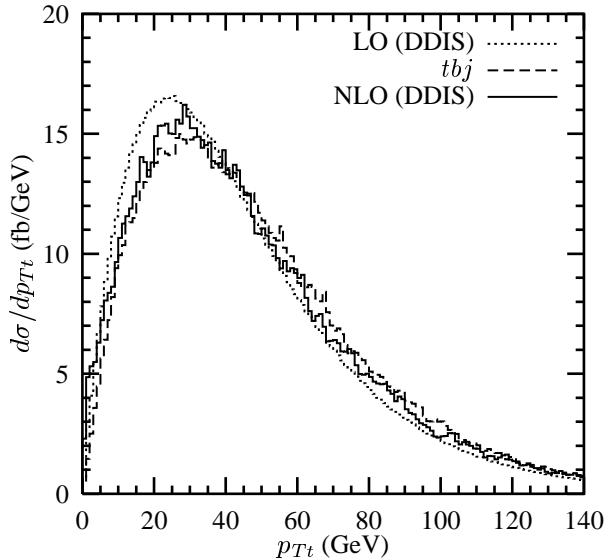}}
\caption{Transverse momentum of the top quark $p_{Tt}$ at LO, NLO, and
from the diagram $tbj$, all normalized to the $t$-channel NLO cross
section.
\label{fig:tbjptt}}
\end{figure}

\begin{figure}[tbh]
\epsfxsize=3.125in 
\centerline{\epsfbox{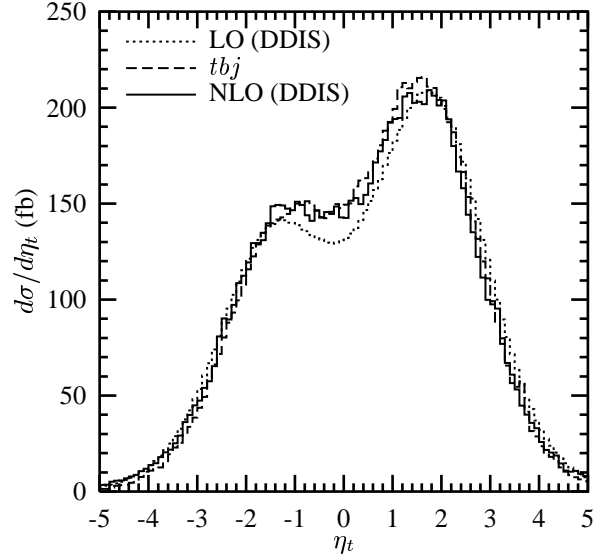}}
\caption{Pseudorapidity of the top quark $\eta_t$ at LO, NLO, and
from the diagram $tbj$, all normalized to the $t$-channel NLO cross
section.
\label{fig:tbjetat}}
\end{figure}

Even though the leading-order diagram for $t$-channel single-top-quark
production was known to include a $b$ quark in the initial state, the
process was modeled by using the $t\bar b j$ diagram shown in
Fig.~\ref{fig:prodt}(b) at leading order in
Refs.~\cite{Stelzer:1998ni,Belyaev:1998dn}.  To the extent that the
analytic form of the cross section is expected to look like double
deep-inelastic-scattering \cite{Stelzer:1997ns,Harris:2002md}, this
diagram was expected to give a reasonable approximation of the shape
of the NLO top-quark distributions.  Indeed, the $p_{Tt}$ and
pseudorapidity $\eta_t$ distributions of the top quark are well
modeled by the LO $t\bar bj$ calculation.  These distributions for $t$
production at the Tevatron (a 1.96 TeV $p\bar p$ collider) are shown
in Figs.~\ref{fig:tbjptt} and \ref{fig:tbjetat}.

\begin{figure}[tbh]
\epsfxsize=3.125in 
\centerline{\epsfbox{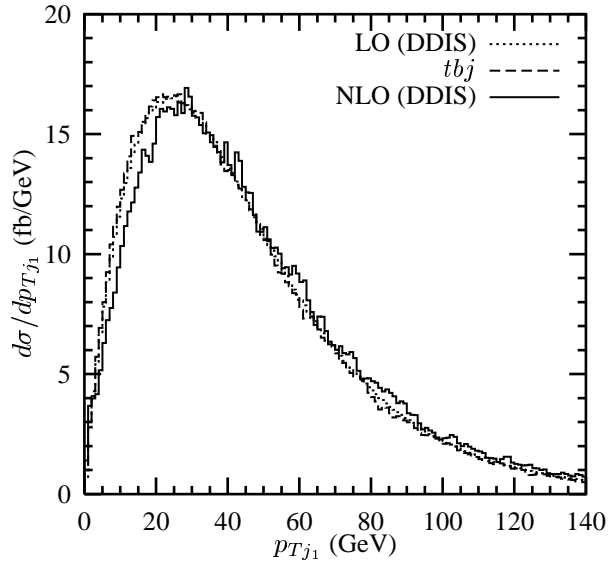}}
\caption{Transverse momentum of the highest-$p_T$ jet $p_{Tj_1}$ at LO,
NLO, and from the diagram $tbj$, all normalized to the $t$-channel NLO
cross section.
\label{fig:tbjptj}}
\end{figure}

\begin{figure}[tbh]
\epsfxsize=3.125in 
\centerline{\epsfbox{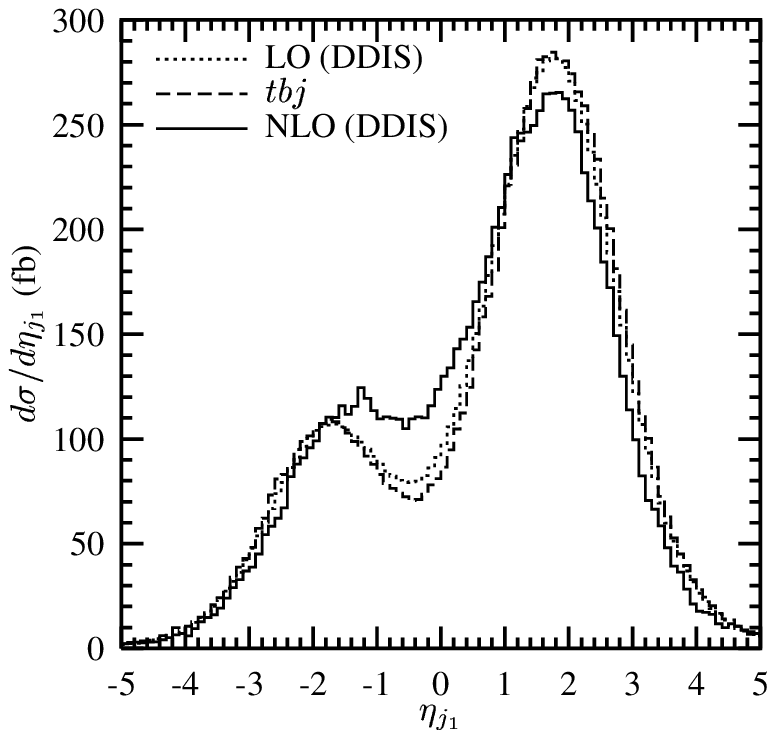}}
\caption{Pseudorapidity of the highest-$p_T$ jet $\eta_{j_1}$ at LO, NLO,
and from the diagram $tbj$, all normalized to the $t$-channel NLO cross
section.
\label{fig:tbjetaj}}
\end{figure}

The diagram $t\bar b j$ should have no visible effect on the
distributions of the highest-$p_T$ jet.  In Figs.~\ref{fig:tbjptj} and
\ref{fig:tbjetaj} we see that there is no difference between the true
LO distributions, and the one obtained from $t\bar b j$ alone.  This
is an additional confirmation that there is a real separation of
physics between the $Wtq$ vertex, and the vertex with no top quark.
While this is a nice confirmation of the double-DIS interpretation,
this means that the distributions for the highest-$p_T$ jet are more
central, and slightly harder than indicated by the old approximation.
Hence analyses, like that in Ref.~\cite{Stelzer:1998ni} slightly
underestimated the true signal.

Finally, let us examine the jet distributions as a first step in
understanding the necessity of the next-to-leading-order calculations.
In particular, we focus on the distributions that will be required for
Sec.~\ref{sec:eventgen}.  The simple cuts in Table~\ref{tab:cuts}
separate the signal into two subsamples: $t+1$ jet exclusive ($Wjj$ in
the experiments, or the ``jet veto'' of Ref.~\cite{Stelzer:1998ni}),
and $t+2$ jet inclusive ($Wjjj$ in the experiments).  Jets are
clustered with a $k_T$ algorithm using $\Delta R_{k_T} < 1.0$ (similar
to a fixed cone of $0.7$) and are observed if they have $p_{Tj}> 15$
GeV and $|\eta_j| < 2.5$.  No cuts are placed on the top quark since
its decay is not modeled.

\begin{table}[tbh]
\caption{Cuts on the $p_T$-ordered jets that define each signal region for
comparison at the Tevatron.  No cuts are placed on the top quark since it
is not decayed.
\label{tab:cuts}}
\medskip
\begin{tabular}{l}\hline\hline
$|\Delta R_{k_T}| < 1.0$ $(\approx |\Delta R_{\mathrm{cone}}| < 0.74)$\\
$p_{T j}>15$ GeV, $|\eta_{j}| < 2.5$\\ \hline
$Wjj\equiv t+1$ jet (``jet veto'')\\
$Wjjj\equiv t+2$ jets inclusive\\ \hline\hline
\end{tabular}
\end{table}

\begin{figure}[tbh]
\epsfxsize=3.125in 
\centerline{\epsfbox{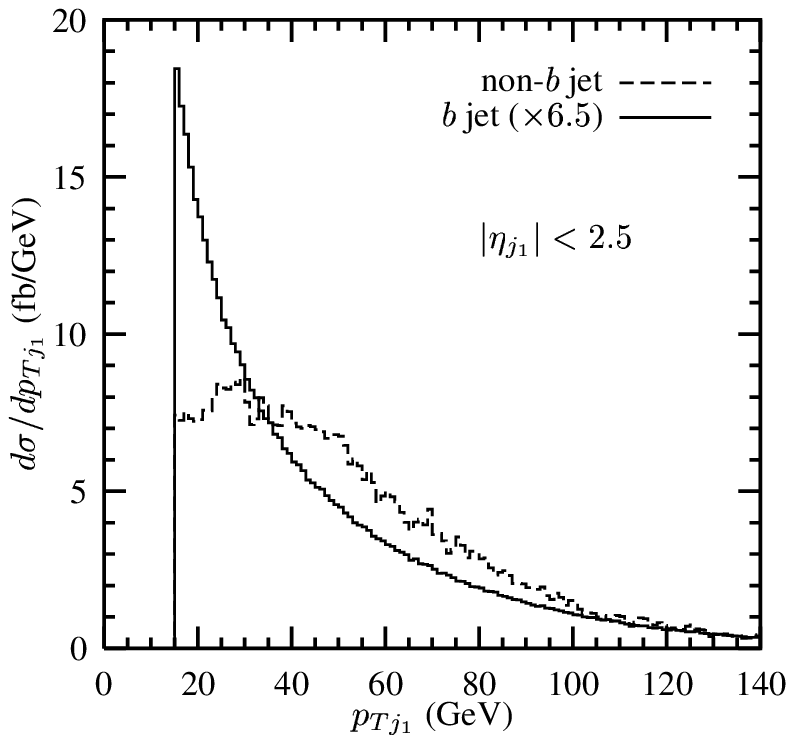}}
\caption{Transverse momentum of the highest-$p_T$ jet $p_{Tj_1}$ in
$t$-channel production when it contains a $b$, normalized to the rate
for when it is a non-$b$ jet in the $tj$ final state.
\label{fig:cptjot}}
\end{figure}

\begin{figure}[tbh]
\epsfxsize=3.125in 
\centerline{\epsfbox{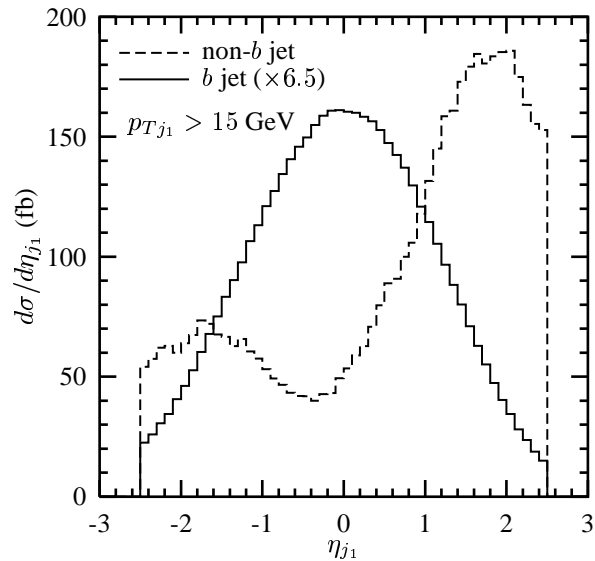}}
\caption{Pseudorapidity of the highest-$p_T$ jet $\eta_{j_1}$ in
$t$-channel production when it contains a $b$, normalized to the rate
for when it is a non-$b$ jet in the $tj$ final state.
\label{fig:cetajot}}
\end{figure}

In Figs.~\ref{fig:cptjot} and \ref{fig:cetajot} we see the transverse
momentum and pseudorapidity distributions for the NLO jet in the $t+1$
jet final state, distinguished by whether there is a $b$ inside the
jet.  The distinction must be made, because the jet containing a $b$
is both potentially taggable, and has a significantly different shape
in the distributions.  Notice that non-$b$ jets exhibit a double-peak
structure, sometimes referred to as a ``forward'' jet.  Whereas, the
additional $b$ jets tend to be very central.

\begin{figure}[tbh]
\epsfxsize=3.125in 
\centerline{\epsfbox{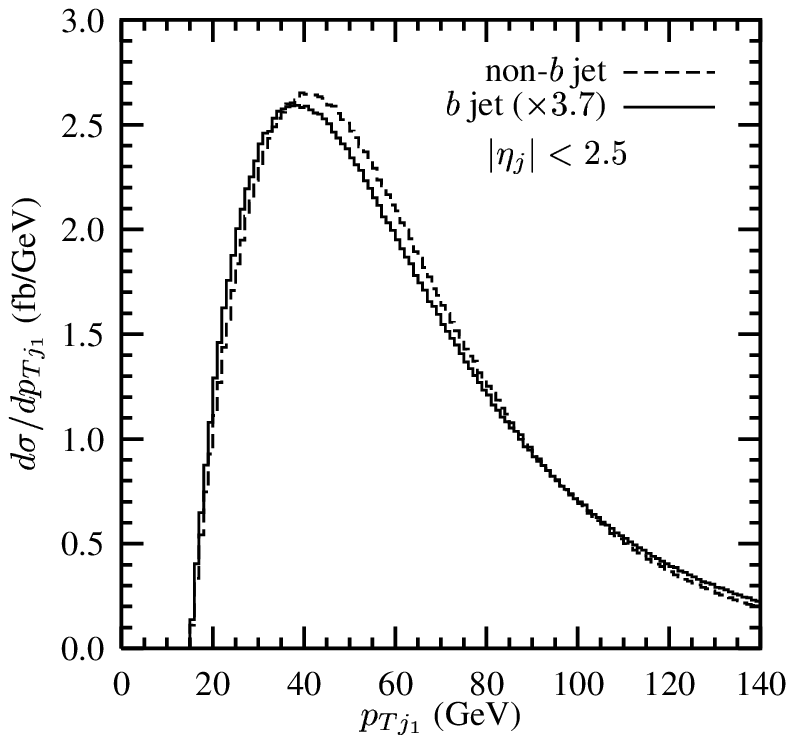}}
\caption{Transverse momentum of the highest-$p_T$ jet $p_{Tj_1}$ in
$t$-channel production when it contains a $b$, normalized to the rate
for when it is a non-$b$ jet in the $tjj$ final state.
\label{fig:cptjott}}
\end{figure}

\begin{figure}[tbh]
\epsfxsize=3.125in 
\centerline{\epsfbox{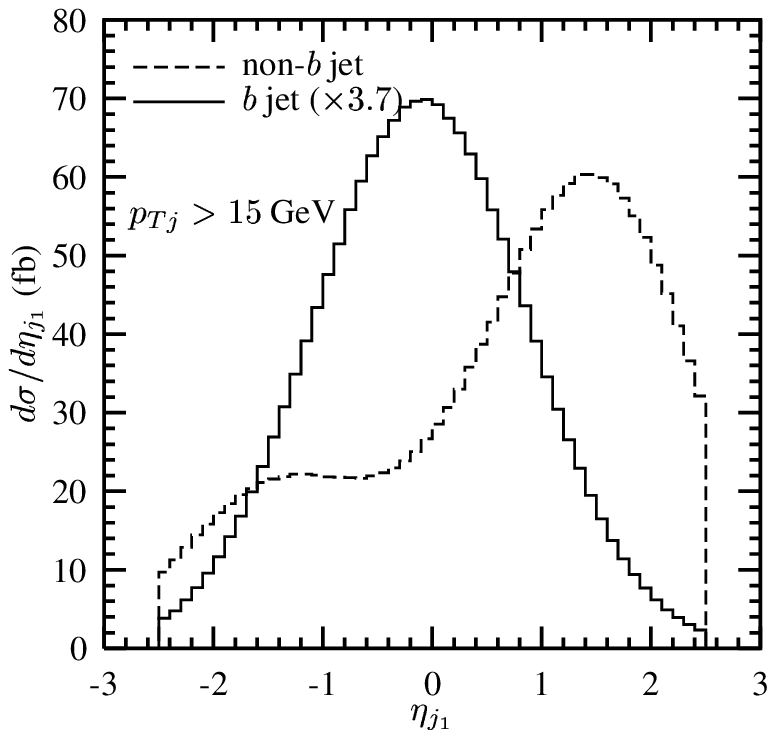}}
\caption{Pseudorapidity of the highest-$p_T$ jet $\eta_{j_1}$ in
$t$-channel production when it contains a $b$, normalized to the rate
for when it is a non-$b$ jet in the $tjj$ final state.
\label{fig:cetajott}}
\end{figure}

\begin{figure}[tbh]
\epsfxsize=3.125in 
\centerline{\epsfbox{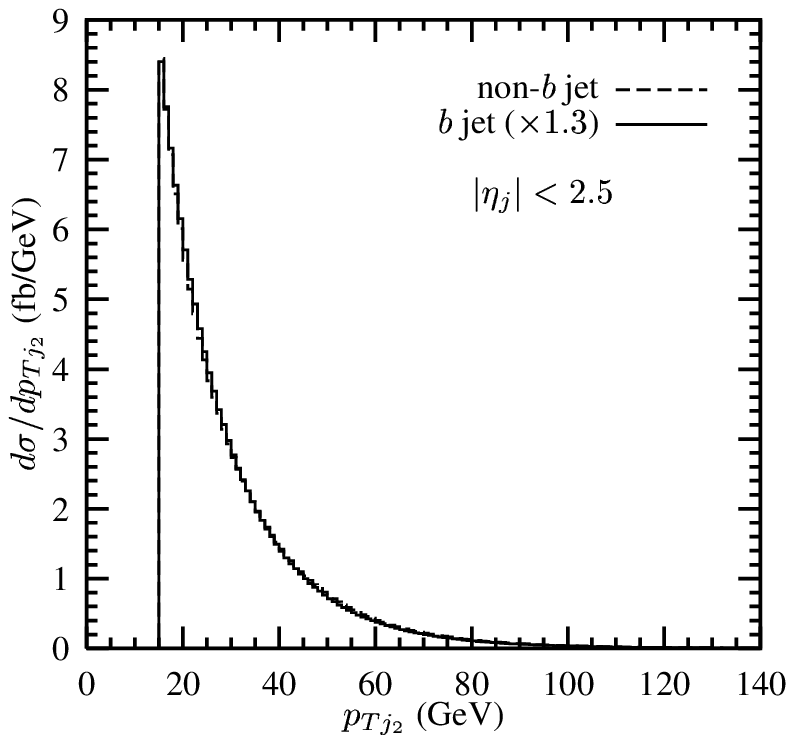}}
\caption{Transverse momentum of the second highest-$p_T$ jet $p_{Tj_2}$
in $t$-channel production when it contains a $b$, normalized to the
rate for when it is a non-$b$ jet in the $tjj$ final state.
\label{fig:cptjttt}}
\end{figure}

\begin{figure}[tbh]
\epsfxsize=3.125in 
\centerline{\epsfbox{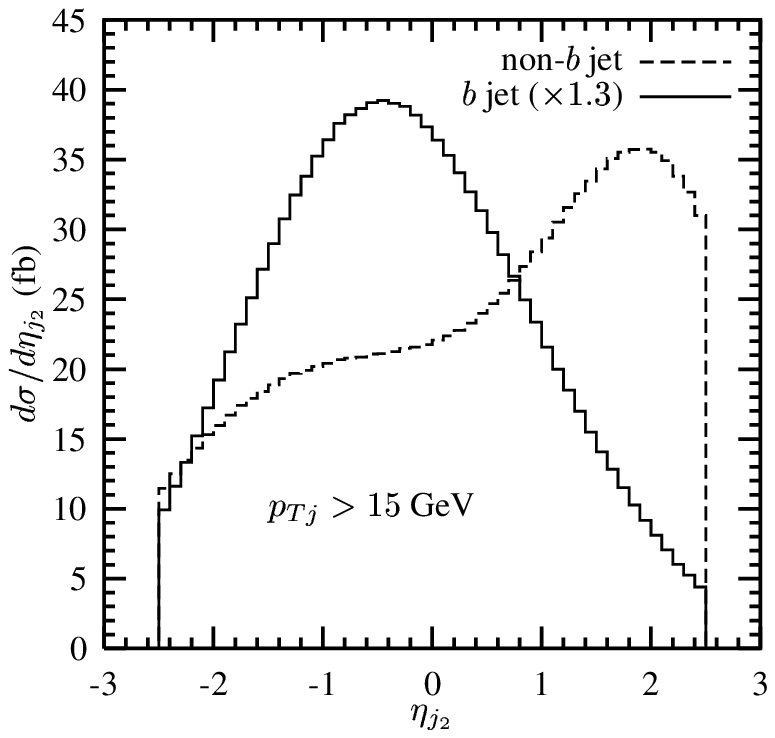}}
\caption{Pseudorapidity of the second highest-$p_T$ jet $\eta_{j_2}$ in
$t$-channel production when it contains a $b$, normalized to the rate
for when it is a non-$b$ jet in the $tjj$ final state.
\label{fig:cetajttt}}
\end{figure}

In Figs.~\ref{fig:cptjott}, \ref{fig:cetajott}, \ref{fig:cptjttt}, and
\ref{fig:cetajttt} we see the distributions for jets in the $t+2$ jet
final state.  The distinction between $t$ plus one and two jet samples
is strongly dependent on the cuts.  Just as in the case of the one jet
sample, the $b$ jets are significantly more central.  However, they
actually peak in the opposite hemisphere from the non-$b$ jets.
Approximating the additional radiation with just a $b$ jet, as in
$t\bar bj$, greatly exaggerates the centrality of the events.  Hence,
the only way to correctly model the signal is to include all
production diagrams in a complete next-to-leading-order calculation.

In theoretical studies that used the $t\bar bj$ diagram to approximate
the final state, all additional radiation was implicitly assumed to
come from an extra $b$ jet.  However, the NLO calculation tells us
that only about $2/3$ of the additional radiation includes a $b$ jet.
Fig.~\ref{fig:cetajot} demonstrates that the contamination from
additional $b$ jets in the $tj$ signal is highly dependent on the
choice of cuts.  For the loose cuts shown, about 15\% of the events
have an extra $b$ jet, but tighter cuts on the pseudorapidity will
greatly enhance the relative contribution from events with an extra
$b$ jet.

The next-to-leading-order calculation is necessary to properly account
for the absolute cross section in each of the four final states: $tj$,
$tb$, $tjj$, and $tbj$.  In Table~\ref{tab:jetpercent} we see the
contribution to each final state for the simple cuts given in Table
\ref{tab:cuts}.  The $b$ jet is the highest-$p_T$ jet in 1/3 of the
events in the $tbj$ final state.  The relative cross sections are
completely insensitive to modest cuts on the top quark itself.
However, it bears repeating that the percentages will change
drastically if the cuts on the jets are changed.  Hence, a careful
matching of the NLO calculation for each choice of cuts is essential
to the proper modeling of these events.

\begin{table}[tbh]
\caption{$t$-channel and $s$-channel cross sections (pb) for $t$ (or
$\bar t$) production at a 1.96 TeV $p\bar p$ collider, and relative
percentages of $b$ jets vs.\ non-$b$ jets $j$ in the $t+1$ jet
exclusive and $t+2$ jet inclusive final states.  These numbers are
only valid for the cuts given in Table \protect\ref{tab:cuts}.
Tighter cuts on pseudorapidity typically increase the percentage of
$b$ jets in the $t$-channel final states.
\label{tab:jetpercent}}
\medskip
\begin{tabular}{cclccl}\hline\hline
$t$-channel & $tj$ & 0.475 pb (87\%) & & $tjj$ & 0.076 pb (35\%) \\
&             $tb$ & 0.073 pb (13\%) & & $tbj$ & 0.139 pb (65\%) \\ \hline
$s$-channel & $tb$ & 0.310 pb (97\%) & & $tbj$ & 0.084 pb \\
&             $tj$ & 0.011 pb (3\%)  & & & \\ \hline\hline
\end{tabular}
\end{table}

\subsection{$s$-channel NLO distributions}
\label{sec:snlo}

\begin{figure}[tbh]
\epsfxsize=3.125in 
\centerline{\epsfbox{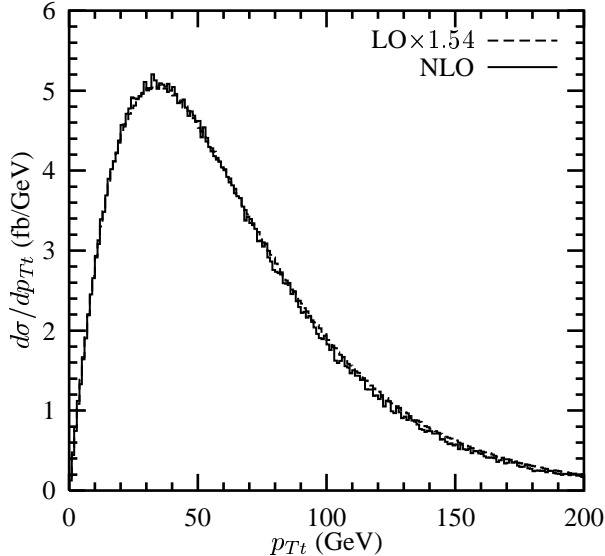}}
\caption{Transverse momentum of the top quark $p_{Tt}$ in $s$-channel
production at NLO, and LO times a $K$-factor of 1.54.
\label{fig:sptt}}
\end{figure}

\begin{figure}[tbh]
\epsfxsize=3.125in 
\centerline{\epsfbox{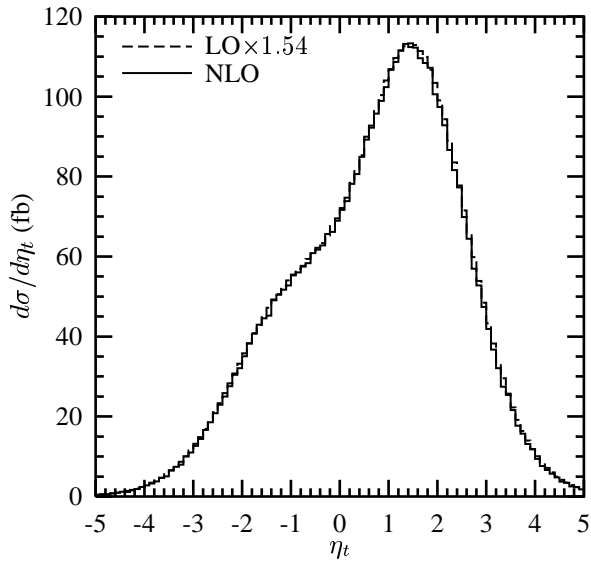}}
\caption{Pseudorapidity of the top quark $\eta_t$ in $s$-channel
production at NLO, and LO times a $K$-factor of 1.54.
\label{fig:setat}}
\end{figure}

\begin{figure}[tbh]
\epsfxsize=3.125in 
\centerline{\epsfbox{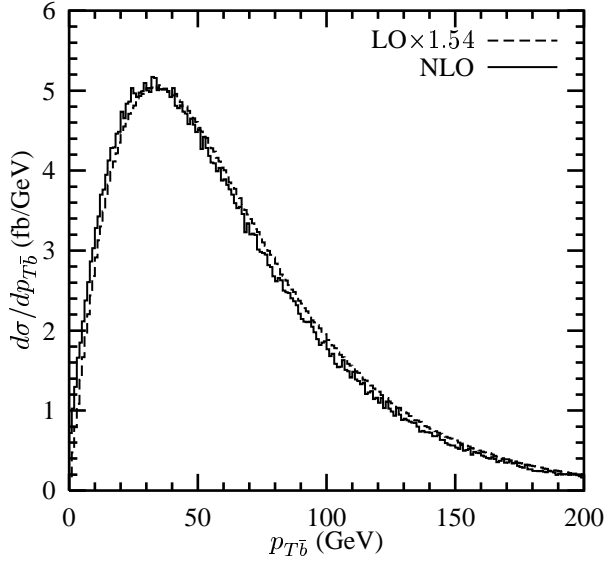}}
\caption{Transverse momentum of the $\bar b$-jet $p_{T \bar b}$ in
$s$-channel production at NLO, and LO times a $K$-factor of 1.54.
\label{fig:sptb}}
\end{figure}

\begin{figure}[tbh]
\epsfxsize=3.125in 
\centerline{\epsfbox{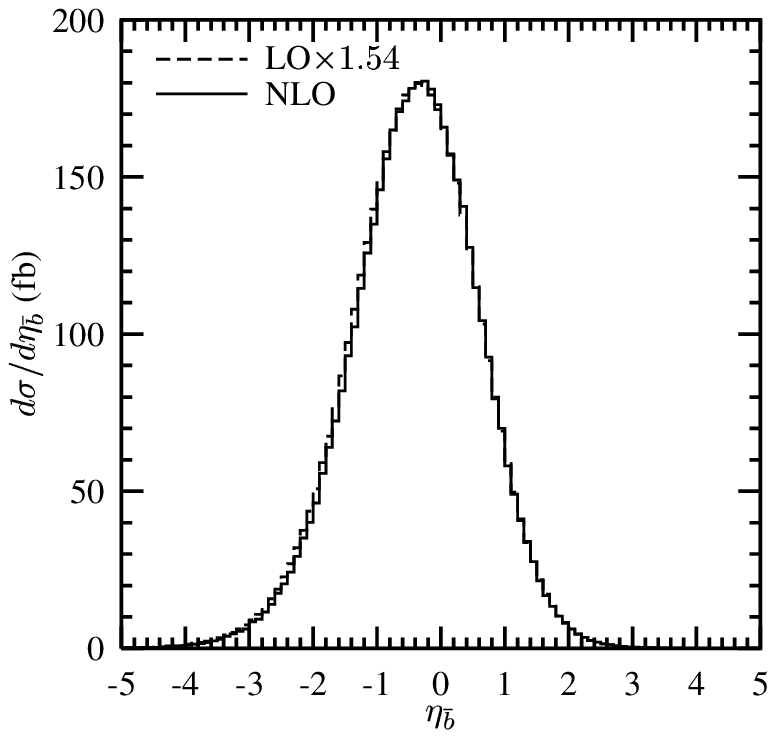}}
\caption{Pseudorapidity of the $\bar b$-jet $\eta_{\bar b}$ in
$s$-channel production at NLO, and LO times a $K$-factor of 1.54.
\label{fig:setab}}
\end{figure}

Production of a single top quark through an $s$-channel $W$ boson
looks very much like Drell-Yan production
\cite{Cortese:fw,Stelzer:1995mi,Dittmar:1997md,Smith:1996ij,Harris:2002md}.
In Figs.~\ref{fig:sptt}, \ref{fig:setat}, \ref{fig:sptb}, and
\ref{fig:setab} we see the transverse momentum $p_T$ and
pseudorapidity $\eta$ of the top quark and $\bar b$ jet.  The large
mass of the top quark protects the distributions from changing shape
at next-to-leading order.  Hence, it is an excellent approximation to
use leading-order distributions and a $K$-factor of 1.54 as long as we
restrict our attention to the $t\bar b$ $(Wb\bar b)$ final state.

Contributions to the $tj$ and $tbj$ final states arise only at
next-to-leading order, and will contribute a small correction to the
total cross section.  In Table~\ref{tab:jetpercent} we see that the
$tj$ channel only contributes 3\% to the $t+1$ jet final state, and
that the ratio between the $t+1$ jet and $t+2$ jet final states is
about 4 to 1.  The separation between the final states is discussed in
detail in Sec.~\ref{sec:genschan}, where we see how the event
generators fare at reproducing the jet distributions.

\subsection{Effect of jet definitions}
\label{sec:jetdef}

Because next-to-leading-order calculations are jet calculations, and
not parton calculations, results are completely dependent on the
choice of jet definition.  This means that no prediction is meaningful
outside of the context of the particular jet definition used.
Historically, jets have been discussed using fixed cones of size
$\Delta R_{\mathrm{cone}}<0.7$, where $\Delta R_{\mathrm{cone}}$ is
the distance between particles in the pseudorapidity $\eta$--azimuthal
angle $\phi$ plane.  This definition suffers from an infrared
instability, because the number of jets can change as particles
fluctuate across the fixed boundary.  One infrared safe
\cite{Ellis:tq} definition is to choose an algorithm that adaptively
combines subjets as a function of their relative transverse momenta
$k_T$.  For this study the $k_T$ cluster algorithm is used with a
merging parameter $\Delta R_{k_T}<1.0$, because it produces jets that
are very similar to those found when using a fixed-cone of size $0.7$.

\begin{figure}[tbh]
\epsfxsize=3.15in 
\centerline{\epsfbox{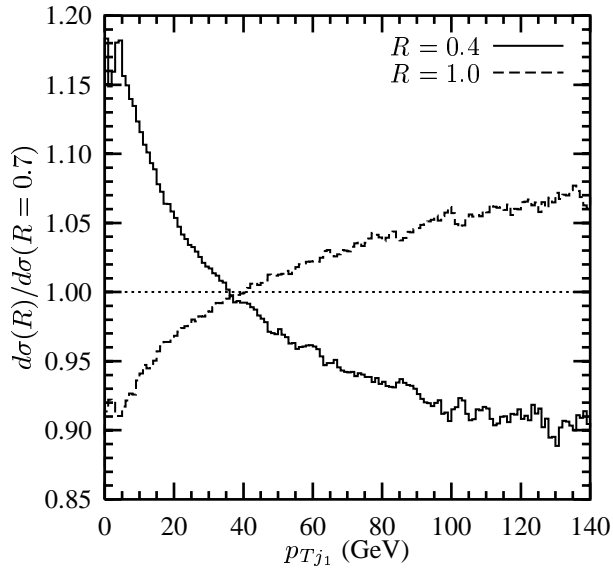}}
\caption{Ratio $d\sigma(R)/dp_{T j_1}$ to $d\sigma(R=0.7)/dp_{T j_1}$
for the highest-$p_T$ jet in $t$-channel production.
\label{fig:delrjo}}
\end{figure}

In principle, the same jet definition must be used for both the theory
and the experimental analyses.  In practice, the NLO calculation will
be matched to an event generator that is tuned to the experimental
data.  The NLO distributions are valid only under the jet definition
used to create them.  If the experiment decides to use smaller cones,
e.g., a cone of size $\Delta R<0.4$ is currently popular for top
physics, the NLO calculation must be reevaluated.  In
Fig.~\ref{fig:delrjo} we see the ratio of the distributions $p_{T
j_1}(\Delta R)$ over $p_{T j_1}(0.7)$, where $j_1$ is the
highest-$p_T$ jet in $t$-channel production, and approximate
fixed-cone units label the curves (the actual $\Delta R_{k_T}$ used
are larger by a factor of 1.35).  We see there is a 10\% change in the
slope of the distribution between different typical cone sizes ($0.4$,
$1.0$) and the standard cone size $0.7$.  A mismatch between the cone
size in the theory calculation and experimental calculation could lead
to a larger error than all theoretical uncertainties combined.

When cuts are made on the leading jet, they are typically made between
15--25 GeV.  From Fig.~\ref{fig:tbjptj} we see that this is close to
the peak of the transverse momentum distribution.  The jets that are
measured have a falling $p_T$ spectrum.  Hence, Fig.~\ref{fig:delrjo}
suggests that it is advisable to use the largest jet size possible.
This would reduce jet-energy corrections for out-of-cone effects, and
increase the signal acceptance, while having little effect on the
$Wjj$ backgrounds.  The $t\bar t$ background will increase slightly,
but is not expected to overcome the increased signal.

\subsection{Effect of bottom-quark mass}
\label{sec:bmass}

\begin{figure}[tbh]
\epsfxsize=3.125in 
\centerline{\epsfbox{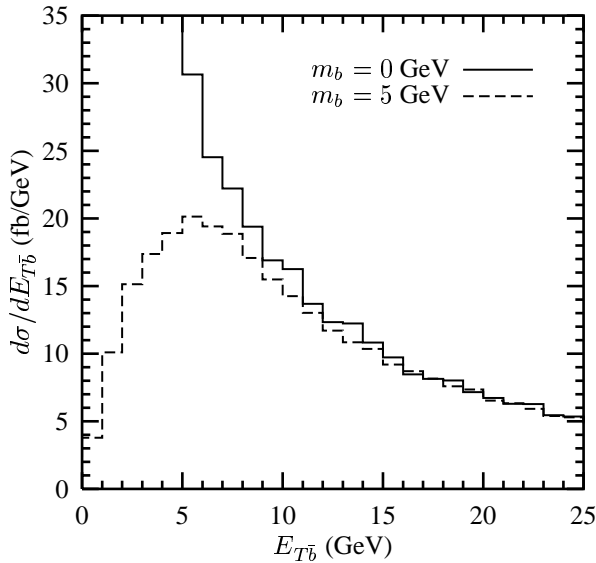}}
\caption{Effect of the $b$-quark mass on the $\bar b$-jet transverse
energy $E_{T\bar b}$ distribution for $t\bar b j$ production with no
cuts.
\label{fig:tptmbnc}}
\end{figure}

When there is an additional bottom quark in the $t$-channel cross
section, we might be concerned whether its mass is relevant to the
calculation.  The same almost-collinear singularity that must be
resummed in the inclusive cross section appears in the distribution of
this additional jet at low transverse momentum.  In
Fig.~\ref{fig:tptmbnc} we see the transverse energy of the extra
bottom jet $E_{T\bar b}$ with no cuts, compared to distribution from
leading-order $t\bar bj$ with a finite $b$ mass.  The curves are very
similar ($<10\%$) above 10 GeV and differ by only a few percent by 15
GeV.  This is confirms the standard belief that a massive particle
behaves the same as a massless particle in the large momentum limit.

\begin{figure}[tbh]
\epsfxsize=3.125in 
\centerline{\epsfbox{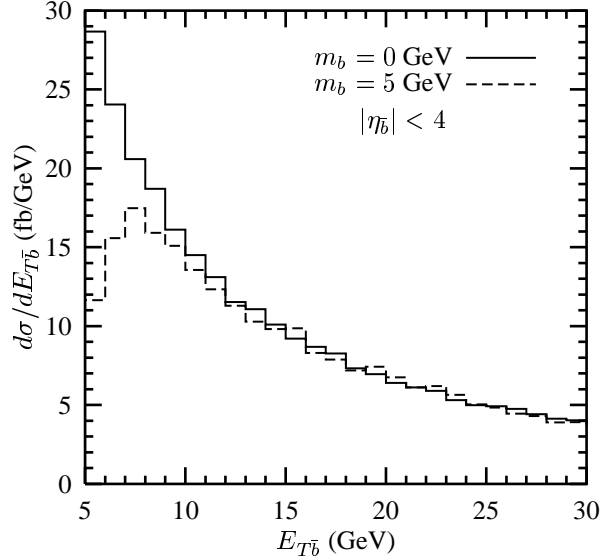}}
\caption{Effect of the $b$-quark mass on the $\bar b$-jet transverse
energy $E_{T\bar b}$ distribution for $t\bar b j$ production with cuts
on both jets $p_{Tj} > 5$ GeV and $|\eta_j| < 4$, and no cut on the
top-quark decay products.
\label{fig:tptmb2j}}
\end{figure}

\begin{figure}[tbh]
\epsfxsize=3.125in 
\centerline{\epsfbox{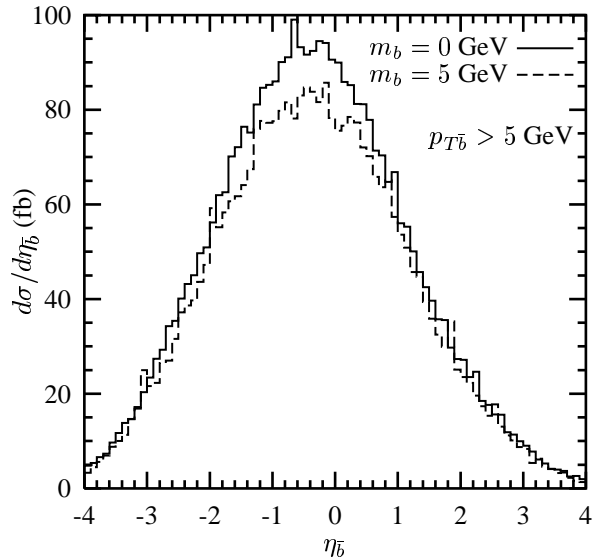}}
\caption{Effect of the $b$-quark mass on the $\bar b$-jet
pseudorapidity $\eta_{\bar b}$ distribution for $t\bar b j$ production
with cuts on both jets $p_{Tj} > 5$ GeV and $|\eta_j| < 4$, and no cut
on the top-quark decay products.
\label{fig:tetamb2j}}
\end{figure}

A more realistic comparison is to consider the difference in jet
distributions after cuts.  The $\bar b$ jet with and without the
bottom-quark mass included is presented in Figs.~\ref{fig:tptmb2j} and
\ref{fig:tetamb2j}, where a minimal set of cuts for potentially
observable jets is defined to have $p_{Tj} > 5$ GeV, and $|\eta_j| <
4$.  In this more realistic case, there is very little difference
between including the bottom quark mass or not.  In any situation
where the $b$-jets are expected to be taggable, the $b$-quark mass
will have less than a 1\% effect on the measured distributions.

\begin{figure}[tbh]
\epsfxsize=3.125in 
\centerline{\epsfbox{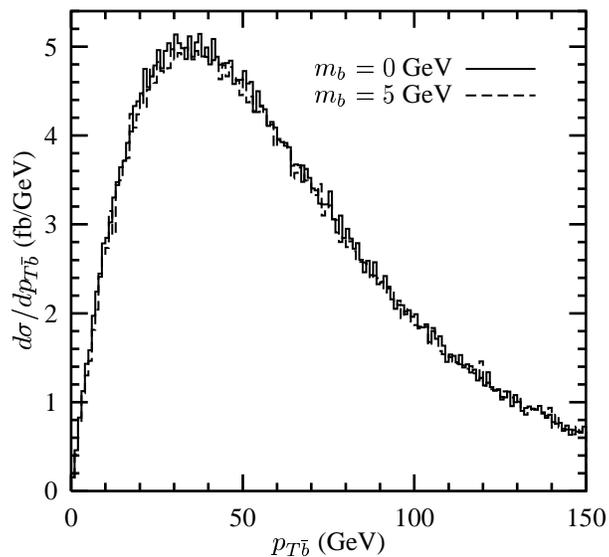}}
\caption{Effect of the $b$-quark mass on the $\bar b$-jet transverse
momentum distribution for $s$-channel single-top-quark production with
no cuts.
\label{fig:sptmbnc}}
\end{figure}

\begin{figure}[tbh]
\epsfxsize=3.125in 
\centerline{\epsfbox{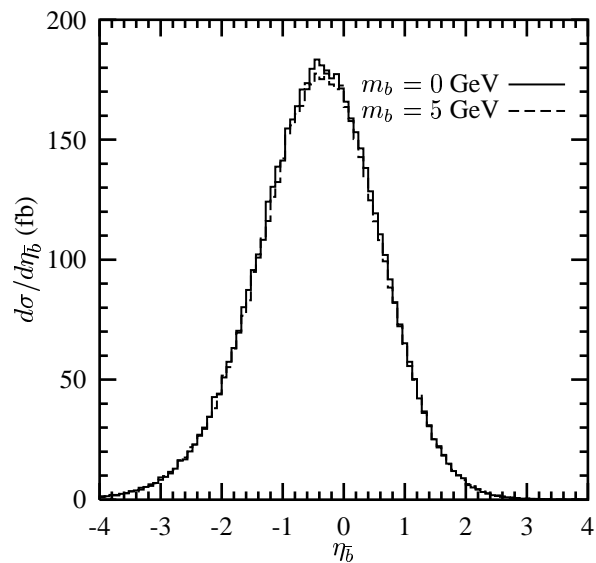}}
\caption{Effect of the $b$-quark mass on the $\bar b$-jet pseudorapidity
distribution for $s$-channel single-top-quark production with no cuts.
\label{fig:setambnc}}
\end{figure}

In the $s$-channel cross section, the extra bottom jet is produced in
the final state, and is protected from any singularities already at LO
by its recoil against the heavy top quark.  We see in
Figs.~\ref{fig:sptmbnc} and \ref{fig:setambnc} that the massless and
massive $b$ distributions are already indistinguishable without any
cuts.  Since there is no visible effect in the measurable portion of
the distributions, we have ignored the bottom-quark mass throughout.
If, in the future, it becomes desirable to measure the third jet below
10 GeV, then mass effects should be taken into account.

\section{Uncertainties}
\label{sec:uncertainties}

Because the measurement of $|V_{tb}|^2$ is directly proportional to
the measurement of the cross section, we must understand the
uncertainties in the extracted cross section.  Historically,
theoretical uncertainties have been expected to dominate the cross
section measurement.  In particular there are two significant sources
of uncertainty, the uncertainty in the quark-gluon luminosity hidden
inside the $b$ parton distribution function (PDF), and the modeling of
the distributions of a third jet that would cause the event to be
vetoed.

In the past, the PDF uncertainty for $t$($s$)-channel single-top-quark
production was approximated by assuming that it was the same as the
gluon(quark) uncertainty.  This crude approximation was shown to
underestimate the uncertainty in the $t$-channel case by a factor of
1.5 \cite{Sullivan:2001ry}.  This study uses the ``Modified Tolerance
Method'' introduced in
Refs.~\cite{Nadolsky:2001yg,Sullivan:2001ry,Sullivan:2002jt} and
CTEQ6M PDFs \cite{Pumplin:2002vw} to provide the first complete
estimation for both $s$-channel and $t$-channel processes.  Results
appear in Table~\ref{tab:uncert}.  The uncertainty at the LHC is much
smaller than expected.  This is partially attributable to the fact
that we are probing the best measured region of the gluon PDFs, which
in turn lowers the $b$ uncertainty.  There is also an anti-correlation
in the quark-$b$ luminosity that stabilizes the calculation.  A more
conservative estimate would use just the $b$ PDF uncertainty of
$^{+3}_{-4.2}\%$.  Newer CTEQ6M1 PDFs \cite{Stump:2003yu} harden the
gluon and sea quarks at moderate fraction of proton momentum $x$.
This gives a $t$-channel cross section for $t$ production of 1.015 pb
(a 2.5\% increase) at the Tevatron, which is well within the current
uncertainty estimate.  In all cases, there is no change in the shapes
of the kinematic distributions.

The shapes of all distributions at the Tevatron and LHC are completely
insensitive to changes of top-quark mass within a few sigma of its
currently measured value.  Nevertheless, the recent average top-quark
mass from run I data of $m_t=178.0\pm 4.3$ GeV \cite{Azzi:2004rc}
reduces the overall normalization of the $s$-channel cross section for
$t$ (or $\bar t$) production used in this paper by 7\% to 0.41 pb at a
1.96 TeV $p\bar p$ collider, and the $t$-channel cross section by
4.8\% to 0.94 pb.  At the LHC (a 14 TeV $pp$ collider), the
$s$-channel cross sections at 178 GeV are 6.17 pb and 3.85 pb, for $t$
and $\bar t$ production, respectively.  The $t$-channel cross sections
are 152.5 pb and 88.7 pb at the LHC, for $t$ and $\bar t$ production,
respectively.  Data from run II of the Tevatron should reduce the
top-quark mass uncertainty to $\pm 2$--3 GeV, but it will remain the
dominant theoretical uncertainty for $s$-channel production until
$\sim 30$ fb$^{-1}$ of data are accumulated.

The complete list of the theoretical uncertainties in the total cross
sections are presented in Table~\ref{tab:uncert} for the Tevatron and
LHC.  The uncertainty due to choice of scale is extensively discussed
in Ref.~\cite{Harris:2002md}, and has no effect on the shapes of the
distributions.  The mass of the $b$ quark, and the error in
$\alpha_s=0.118\pm 0.003$, play an insignificant role in the overall
uncertainty for measurable quantities.  For convenience, the NLO cross
sections and their current total uncertainties are listed in
Table~\ref{tab:sigtot} using CTEQ5M1 PDFs \cite{Lai:1999wy}.  $t$ and
$\bar t$ cross sections are the same at the Tevatron, but are listed
separately at the LHC since it is a $pp$ collider.

\begin{table}[tbh]
\caption{Uncertainties of the next-to-leading-order cross section at the
Tevatron ($p\bar p$) and LHC ($pp$).
\label{tab:uncert}}
\medskip
\begin{tabular}{lccccc}\hline\hline
Process& $\times \delta m_t$(GeV)& $\mu/2$--$2\mu$ & PDF & $b$ mass & $\alpha_s
(\delta_{\mathrm{NLO}})$ \\ \hline
$s$-channel $p\bar p$ & $^{-2.33}_{+2.71}\%$&
 $^{+5.7}_{-5.0}\%$ & $^{+4.7}_{-3.9}\%$ & $< 0.5\%$ & $\pm 1.4\%$ \\
\phantom{$s$-channel} $p p$ & $^{-1.97}_{+2.26}\%$ & $\pm 2\%$ &
 $^{+3.3}_{-3.9}\%$ & $< 0.4\%$ & $\pm 1.2\%$ \\ \hline
$t$-channel $p\bar p$ & $^{-1.6}_{+1.75}\%$ & $\pm 4\%$ &
 $^{+11.3}_{-8.1}\%$  & $<1\%$ & $\pm 0.01\%$ \\
\phantom{$t$-channel} $p p$ & $^{-0.73}_{+0.78}\%$ & $\pm 3\%$ &
 $^{+1.3}_{-2.2}\%$ & $<1\%$ & $\pm 0.1\%$ \\ \hline\hline
\end{tabular}
\end{table}

\begin{table}[tb]
\caption{Next-to-leading-order inclusive cross sections for single-top-quark
production at the Tevatron and LHC for $m_t=175$ GeV.  Cross sections
are evaluated with CTEQ5M1 PDFs, and $M_{t\bar b}$ or the DDIS scales
($\mu_l=Q^2$, $\mu_h=Q^2 + m_t^2$), for $s$-channel or $t$-channel,
respectively.  Theoretical uncertainties are added in quadrature, and
are based on an uncertainty in the top-quark mass of $4.3$
GeV.\label{tab:sigtot}}
\medskip
\begin{tabular}{c|r@{~}l|d@{~}l} \hline\hline
Process&\multicolumn{2}{c|}{$\sqrt{S}$}&
\multicolumn{2}{c}{$\sigma_{\mathrm{NLO}}$ (pb)} \\
\hline
$s$-channel&1.8& TeV $p\bar p$ ($t$) & 0.377 & $^{+0.052}_{-0.045}$ \\
&1.96& TeV $p\bar p$ ($t$) & 0.442 & $^{+0.061}_{-0.053}$ \\
&14& TeV $pp$ ($t$) & 6.56 & $^{+0.69}_{-0.63}$ \\
&14& TeV $pp$ ($\bar t $) & 4.09 & $^{+0.43}_{-0.39}$ \\
\hline
$t$-channel&1.8& TeV $p\bar p$ ($t$) & 0.73 & $^{+0.10}_{-0.08}$ \\
&1.96& TeV $p\bar p$ ($t$) & 0.99 & $^{+0.14}_{-0.11}$ \\
&14& TeV $pp$ ($t$) & 155.9 & $^{+7.5}_{-7.7}$ \\
&14& TeV $pp$ ($\bar t $) & 90.7 & $^{+4.3}_{-4.5}$ \\
\hline\hline
\end{tabular}
\end{table}

Once we begin to calculate exclusive final states, there are
additional uncertainties that arise from modeling the kinematic
distributions.  In Secs.~\ref{sec:etvsnlo} and \ref{sec:snlo} we saw
that NLO distributions of the top quark and leading jet were similar
to LO distributions including $K$-factors if physically motivated
scales were chosen.  The main uncertainty arises in how well the
separation between $t+1$ jet and $t+2$ jet distributions is modeled.
Since the NLO shapes of the jets are stable, the remaining uncertainty
is entirely due to the cuts on the $t+2$ jet final states, which are
only evaluated at LO.

Typical uncertainties may be derived by observing the change in the
normalization of the jet distributions when the scales are varied by a
factor of 2.  The uncertainties for each final state depend on the
exact cuts that are used.  In Table~\ref{tab:derror} I list the
uncertainties for the cuts in Table~\protect\ref{tab:cuts} and cross
sections in Table~\ref{tab:jetpercent}.  I also provide the complete
percentage uncertainty in parentheses, assuming $\delta m_t=4.3$ GeV,
and the uncertainties in Table~\ref{tab:uncert}.  The uncertainty for
a final state containing a top quark plus one or two jets is presented
first.  The subsamples containing zero or one taggable $b$ jets are
also listed.

For $s$-channel production, the $tb$ final state is well modeled, and
the uncertainty is completely dominated by the top-quark mass
measurement.  The kinematic uncertainty is less than 2.5\%.  The $tj$
and $tbj$ final states are LO cross sections, and so are more
sensitive to fluctuations of the additional radiation.  The $tj$ and
$tbj$ cross sections have approximately $^{+20}_{-16}\%$ kinematic
uncertainties in their normalizations, but have very low rates, and
are not necessary for observation of the signal.

In the $t$-channel, the uncertainty in the $tj$ final state was
previously estimated \cite{Stelzer:1998ni} to be $\pm$15\%, based
solely on how often the $t\bar bj$ Feynman diagram produced a $b$ jet
with $p_{Tb}< 15$ GeV.  The same large logarithms that lead to the
need for the introduction of a $b$ PDF, also produce a large
uncertainty in the shape and normalization of the $b$ jet
distributions.  Changes in the two scales in the problem ($Q^2$ and
$Q^2+m_t^2$) tend to accidentally cancel.  Therefore, an accurate
estimation of uncertainty requires varying the scales independently.

The full NLO calculation tells us that the kinematic uncertainty of
the jet in the $tj$ final state is only about 6.5\% if flavor is
ignored.  The decrease comes from two effects.  First, only $2/3$ of
the additional radiation that is cut on includes the poorly modeled
$b$ jet distribution.  Second, the addition of a cut in pseudorapidity
removes the forward $b$ jets that are missed.  Hence, the kinematic
uncertainty in the $tj$ final state is smaller than either the PDF or
top-quark mass uncertainties.  The NLO distribution of $tj$ when the
jet does not contain a $b$ is known to $^{+7.7}_{-10}\%$.  This is
still better than, but comparable to the PDF and top-quark mass
uncertainties.  If tighter cuts are made, this uncertainty will
improve.  Finally, the LO $tb$ final state still has an uncertainty of
about $^{+16}_{-12}\%$.  In general, more inclusive cross sections are
better estimated.

\begin{table}[tb]
\caption{Uncertainties in normalization of jet distributions (pb) in
the $t+n$ jet final states for $t$ (or $\bar t$) production at a 1.96
TeV $p\bar p$ collider.  These uncertainties are valid for the cuts in
Table~\protect\ref{tab:cuts} and cross sections in
Table~\ref{tab:jetpercent}.  Total percentage uncertainties are listed
in parentheses, and include the uncertainties from
Table~\protect\ref{tab:uncert} added in quadrature assuming $\delta
m_t=4.3$ GeV.\label{tab:derror}}
\medskip
\begin{tabular}{llcc}\hline\hline
Process&$j$ flavors&$t+1$ jet &$t+2$ jets\\
\hline
$t$-channel & all & $_{-0.036}^{+0.028}$ pb $\left(_{-13}^{+15}\%\right)$ &
$_{-0.019}^{+0.025}$ pb $\left(_{-14}^{+18}\%\right)$ \\
            & 0 $b$ jets &
$_{-0.048}^{+0.037}$ pb $\left(_{ -15 }^{+ 16 }\%\right)$ &
$_{-0.014}^{+0.018}$ pb $\left(_{ -14 }^{+ 17 }\%\right)$ \\
            & 1 $b$ jet &
$_{-0.009}^{+0.011}$ pb $\left(_{ -16 }^{+ 21 }\%\right)$ &
$_{-0.005}^{+0.007}$ pb $\left(_{ -16 }^{+ 21 }\%\right)$ \\ \hline
$s$-channel & all & $^{+0.007}_{-0.009}$ pb $\left(_{ -11 }^{+ 13 }\%\right)$ &
$^{+0.017}_{-0.013}$ pb $\left(_{ -19 }^{+ 24 }\%\right)$ \\
            & 1 $b$ jet&
$^{+0.005}_{-0.007}$ pb $\left(_{ -11 }^{+ 13 }\%\right)$ &
$^{+0.013}_{-0.010}$ pb $\left(_{ -19 }^{+ 24 }\%\right)$ \\
            & 0 $b$ jets&
$^{+0.002}_{-0.002}$ pb $\left(_{ -19 }^{+ 24 }\%\right)$
& \\ \hline\hline
\end{tabular}
\end{table}

There are two remaining uncertainties that can appear in the
theoretical calculation, but which are not estimated here.  In
Sec.~\ref{sec:eventgen} a matching between the NLO calculation and
event generators is proposed.  Uncertainties arising from the
normalization of the distributions produced by the event generators
can only be calculated when the fits are performed.  The last
uncertainty comes from the fact that the top quark is not decayed in
this analysis.  A leading-order estimate already limits this
uncertainty to less than 2\% for the distributions and search strategy
discussed here.  However, if angular correlations are measured
\cite{Mahlon:1996pn,Stelzer:1998ni}, or extreme cuts are used, then
this uncertainty may be enhanced.  A simple remedy would be to include
the decay of the top quark by combining the complete published
next-to-leading-order spin-density matrices for production
\cite{Harris:2002md} and decay \cite{Czarnecki:1990pe} into a single
program.

\section{Comparison to event generators}
\label{sec:eventgen}

The motivation for performing differential calculations at
next-to-leading order is to improve the connection between theory and
experimental results.  Typically, this connection is made through the
use of a showering event generator.  These event generators begin with
leading-order matrix elements, and model the effects of soft initial
and final-state showering, and hadronization.  The essential question
is whether these modified leading-order calculations are sufficiently
predictive for the final states that will be measured.  For PYTHIA
\cite{Sjostrand:2000wi} and HERWIG \cite{Corcella:2000bw}, the answer
is a resounding no.

\subsection{$t$-channel production in HERWIG and PYTHIA}
\label{sec:gentchan}

\begin{figure}[tbh]
\epsfxsize=2.625in 
\centerline{\epsfbox{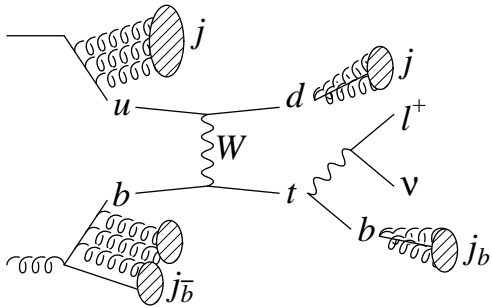}}
\caption{A symbolic representation of $t$-channel single-top-quark production
as generated by HERWIG and PYTHIA.  Soft angular-ordered showering is
applied to the hard matrix element for $ub\to td$.  The radiation is
hadronized, and resummed using a jet definition.
\label{fig:phprod}}
\end{figure}

\begin{figure}[tbh]
\epsfxsize=3.15in 
\centerline{\epsfbox{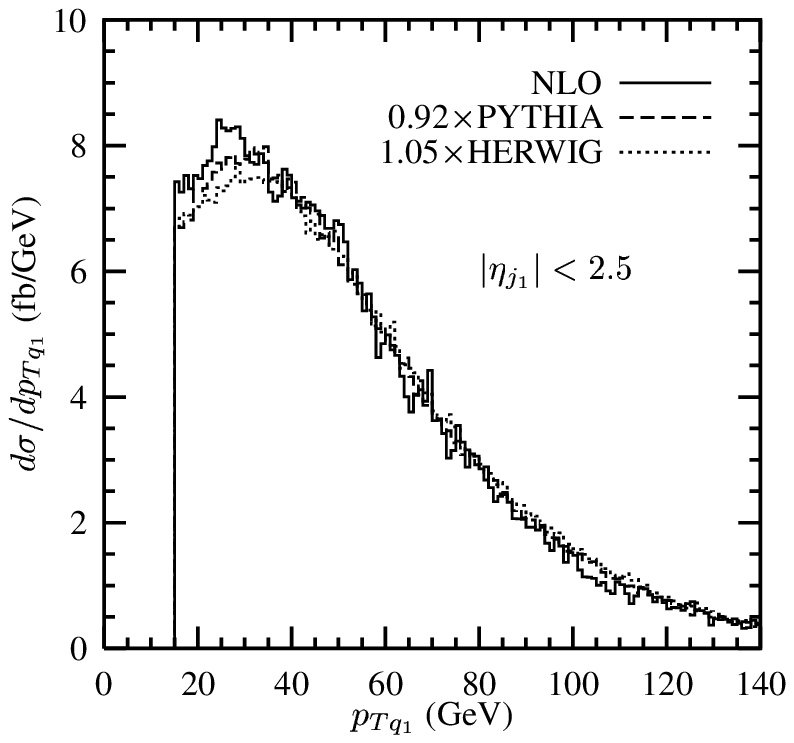}}
\caption{Transverse momentum $p_{T q_1}$ of the observed non-$b$ jet in the
$t+1$ jet final state (``jet veto'') of $t$-channel production from
NLO (solid), PYTHIA (dashed), and HERWIG (short-dashed) normalized to
NLO.
\label{fig:phpqot}}
\end{figure}

\begin{figure}[tbh]
\epsfxsize=3.15in 
\centerline{\epsfbox{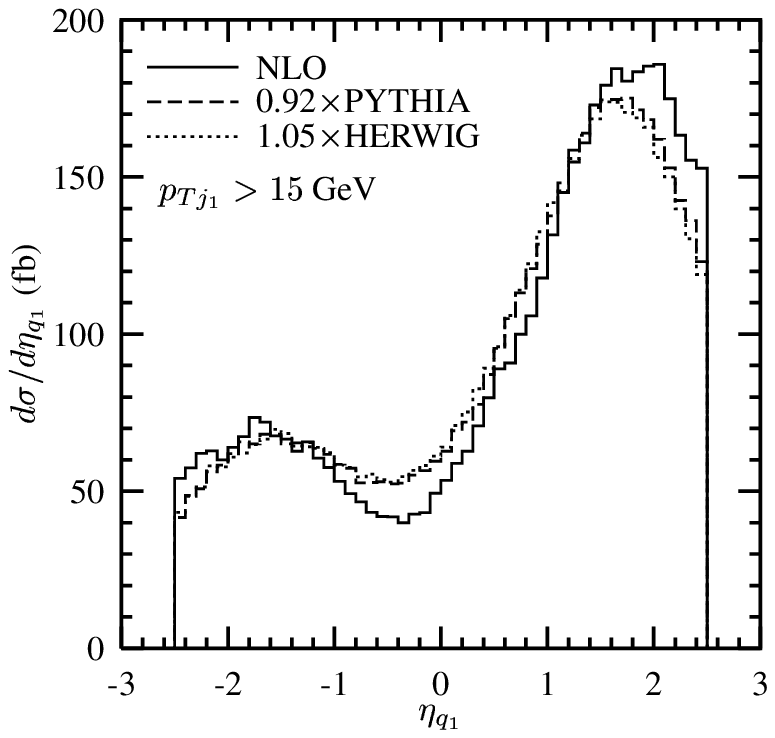}}
\caption{Pseudorapidity $\eta_{q_1}$ of the observed non-$b$ jet in the
$t+1$ jet final state (``jet veto'') of $t$-channel production from
NLO (solid), PYTHIA (dashed), and HERWIG (short-dashed) normalized to
NLO.
\label{fig:pheqot}}
\end{figure}

We begin by examining $t$-channel single-top-quark production at the
Tevatron (see Ref.~\cite{Sullivan:2003xy} for a summary of the
situation at the LHC).  PYTHIA and HERWIG both produce events by
evaluating the hard cross section represented by
Fig.~\ref{fig:phprod}.  Beginning with the correct LO diagram for
$t+1$ jet production, the programs apply soft showering to the
final-state quarks (which does not change the primary direction of the
overall jet), and use strong-angular-ordered showering for incoming
quarks.  Using the cuts of Table~\ref{tab:cuts}, we see in
Figs.~\ref{fig:phpqot} and \ref{fig:pheqot} that the extra jet is
reasonably well represented if it is \emph{not} a $b$ jet.  PYTHIA and
HERWIG have been normalized to the inclusive NLO cross section (the
$K$-factor is the same if you normalize to the $t+1$ jet inclusive
cross section after cuts).  The main discrepancy here is that PYTHIA
overestimates the signal by 8\%, while HERWIG underestimates it by 5\%.

\begin{figure}[tbh]
\epsfxsize=3.15in 
\centerline{\epsfbox{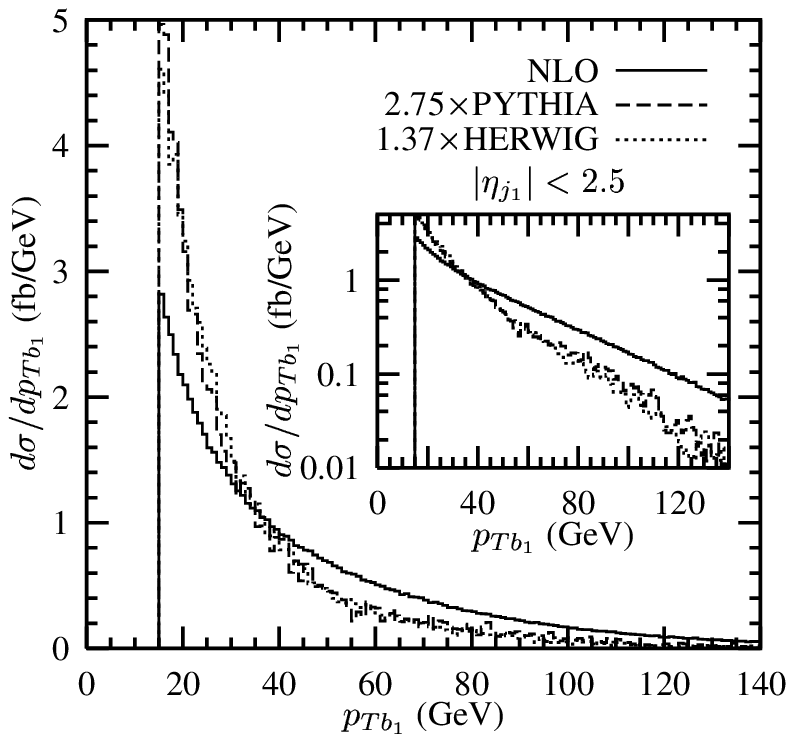}}
\caption{Transverse momentum $p_{T b_1}$ of the observed $b$ jet in the
$t+1$ jet final state (``jet veto'') of $t$-channel production from
NLO (solid), PYTHIA (dashed), and HERWIG (short-dashed) normalized to
NLO.  Inset is on a logarithmic scale.
\label{fig:phpbot}}
\end{figure}

\begin{figure}[tbh]
\epsfxsize=3.15in 
\centerline{\epsfbox{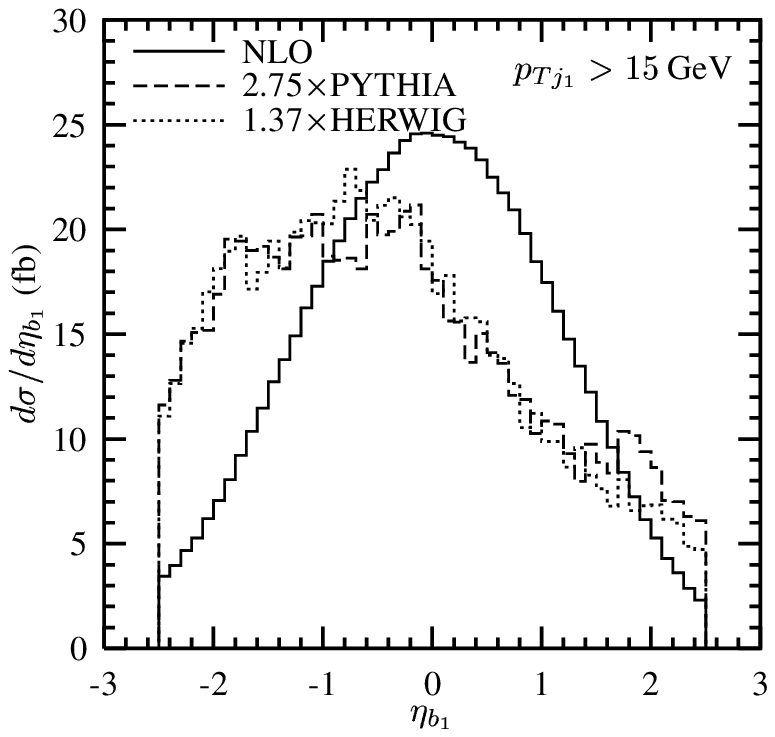}}
\caption{Pseudorapidity $\eta_{b_1}$ of the observed $b$ jet $b_1$ in the
$t+1$ jet final state (``jet veto'') of $t$-channel production from NLO
(solid), PYTHIA (dashed), and HERWIG (short-dashed) normalized to NLO.
\label{fig:phebot}}
\end{figure}

Unfortunately, both PYTHIA and HERWIG predict completely incorrect
distributions for the $b$ jet in the $t+b$-jet sample.  In
Fig.~\ref{fig:phpbot} we see that the transverse momentum predicted by
the event generators is too soft.  While we can imagine trying to
harden the spectrum by tuning parameters of the event generator, we
cannot tune the pseudorapidity distributions of Fig.~\ref{fig:phebot}.
The showering event generators are producing $b$ jets that are too far
forward.  Once central cuts are placed, PYTHIA (HERWIG) underestimates
the number of events in this channel by at least a factor of $2.75$
$(1.4)$; tighter cuts produce a greater underestimate.  The origin of
the problem is that the showering event generators only produce
additional $b$ jets from the soft angular-ordered initial-state
radiation.  All jets produced this way lead to distributions that are
too soft and too far forward.

The failure of PYTHIA and HERWIG to model the $t+b$-jet final state has
several important consequences: Analyses and acceptance studies using
PYTHIA or HERWIG will lead to approximately a factor of 3
underestimate of the cross section in the $Wb\bar b$ bin.  This means
the cross section for discovery is larger than currently estimated.
However, this also means that the background to other physics analyses
has been severely underestimated.  The obvious case is for $s$-channel
single-top-quark production, which is distinguished from $t$-channel
production by the number of $b$ tags.  Also, the Higgs analysis
($WH\to Wb\bar b$) will have a significantly larger contribution from
$t$-channel production than estimated.  For lighter Higgs masses
($<130$ GeV) single-top-quark production may be a more significant
background than $t\bar t$ production.

\begin{figure}[tbh]
\epsfxsize=7in 
\centerline{\epsfbox{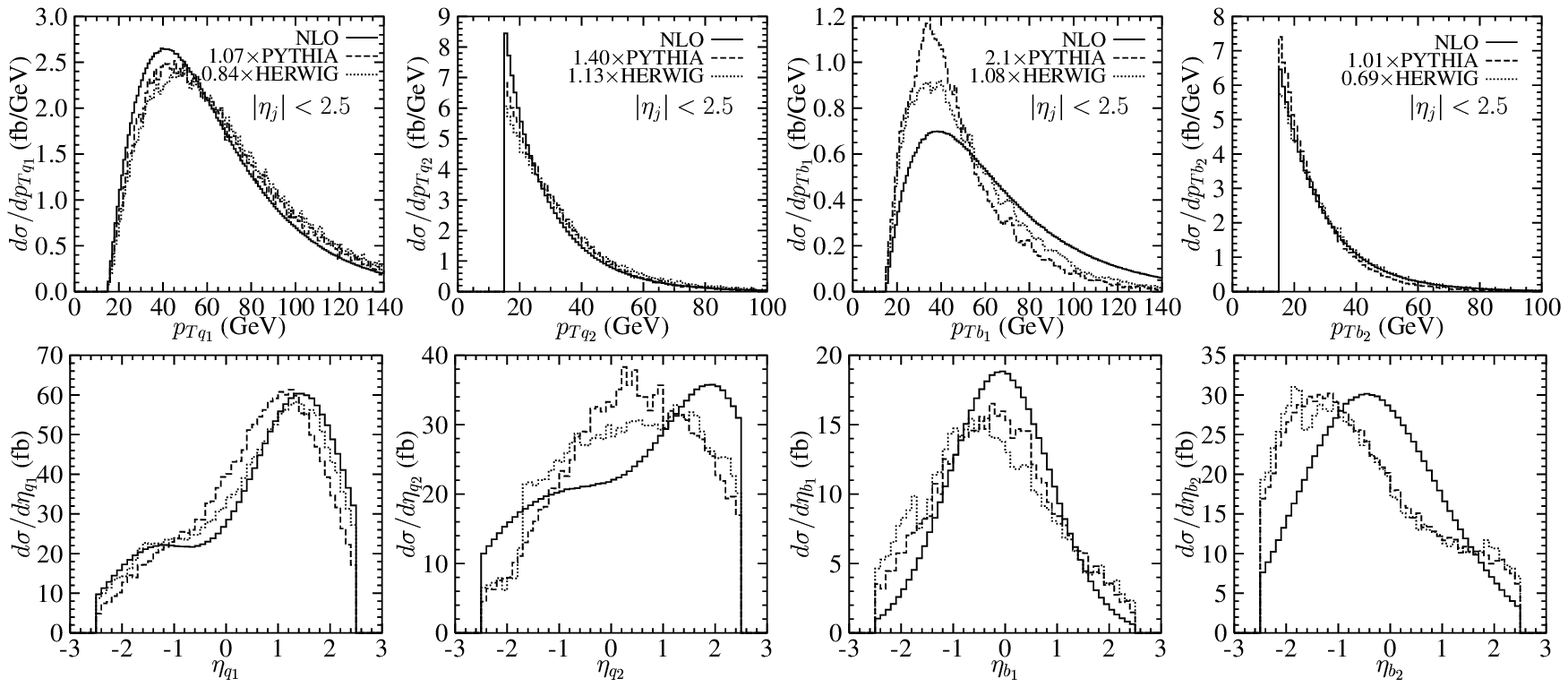}}
\caption{Transverse momentum $p_T$ and pseudorapidity $\eta$ for the
$p_T$-ordered jets $j_i$ in the $t+2$-jet final state of $t$-channel
production from NLO (solid), PYTHIA (dashed), and HERWIG
(short-dashed). $j_1$ and $j_2$ are divided into non-$b$ jets $q_i$
and $b$ jets $b_i$.  Curves are normalized to NLO in each column.
\label{fig:phjt}}
\end{figure}

The shapes of the distributions in the $t+2$-jet final state are also
not well-modeled by HERWIG and PYTHIA.  In Fig.~\ref{fig:phjt} we see
the transverse momentum and pseudorapidities of $b$ and non-$b$ jets
scaled to the NLO event rate.  The non-$b$ leading jets $q_1$ are
reasonable in shape, but either over or underestimated in rate by up
to 15\%.  Other distributions are off in rate by up to a factor of 2,
and the pseudorapidities are poorly reproduced by both HERWIG and
PYTHIA.  A method for correcting these samples is discussed in
Sec.~\ref{sec:gencor}.

\subsection{$s$-channel production in HERWIG and PYTHIA}
\label{sec:genschan}

\begin{figure}[tbh]
\epsfxsize=3in 
\centerline{\epsfbox{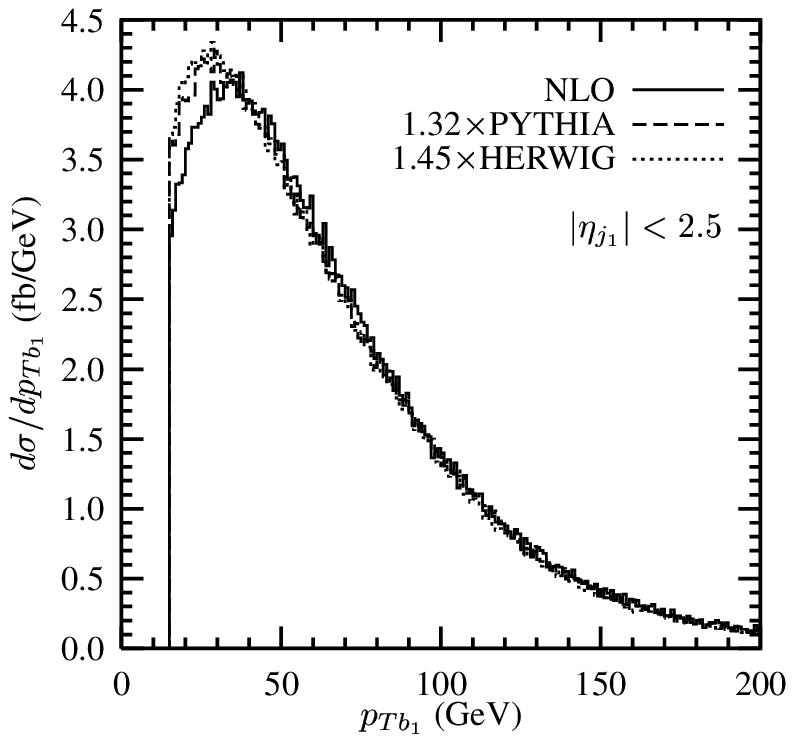}}
\caption{Transverse momentum $p_{T b_1}$ of the observed $b$ jet in the
$t+1$ jet final state (``jet veto'') of $s$-channel production from NLO
(solid), PYTHIA (dashed), and HERWIG (short-dashed) normalized to NLO.
\label{fig:phptbs}}
\end{figure}

\begin{figure}[tbh]
\epsfxsize=3in 
\centerline{\epsfbox{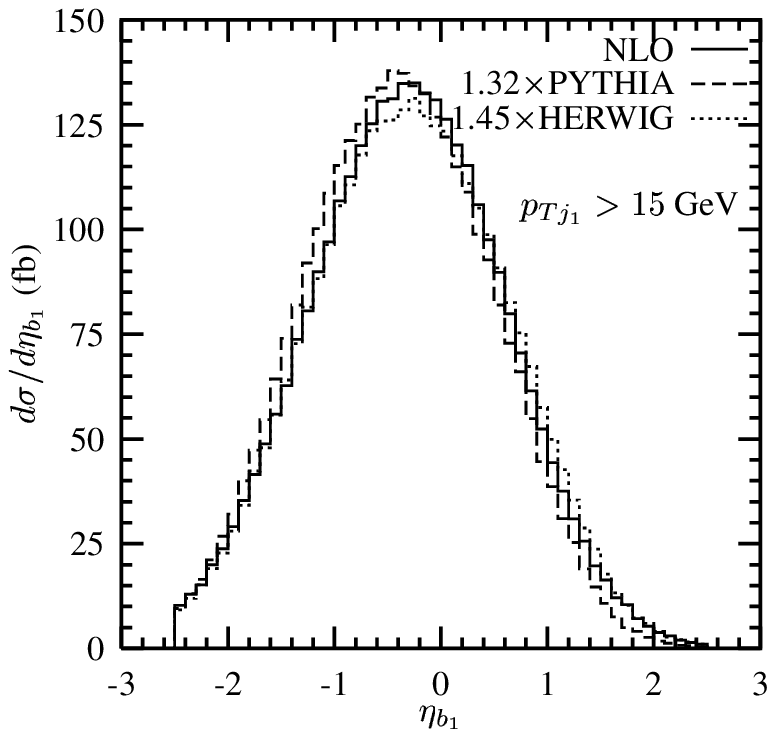}}
\caption{Pseudorapidity $\eta_{b_1}$ of the observed $b$ jet $b_1$ in the
$t+1$ jet final state (``jet veto'') of $s$-channel production from NLO
(solid), PYTHIA (dashed), and HERWIG (short-dashed) normalized to NLO.
\label{fig:phetabs}}
\end{figure}

\begin{figure}[tbh]
\epsfxsize=3in 
\centerline{\epsfbox{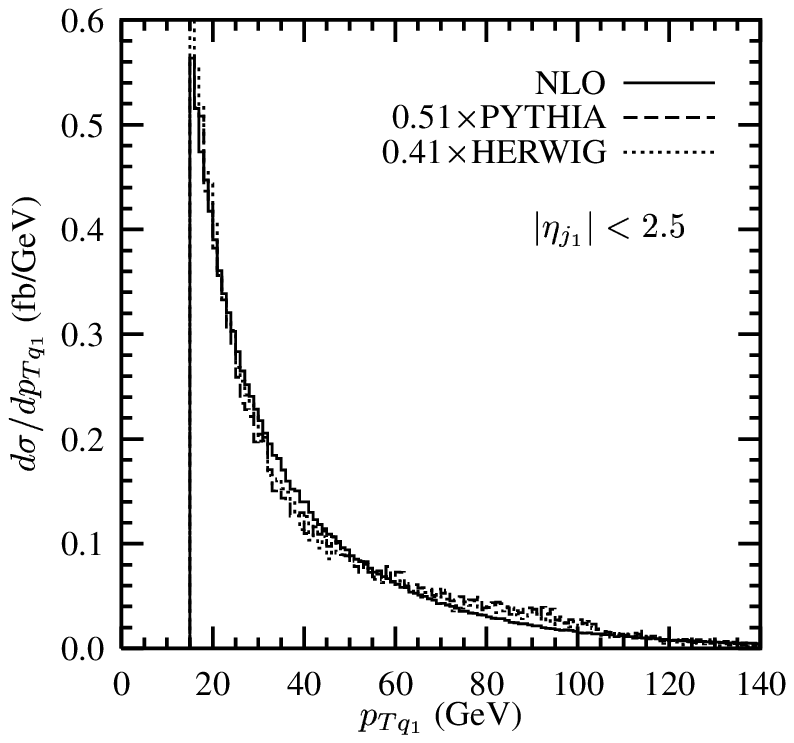}}
\caption{Transverse momentum $p_{T q_1}$ of the observed non-$b$ jet in the
$t+1$ jet final state (``jet veto'') of $s$-channel production from NLO
(solid), PYTHIA (dashed), and HERWIG (short-dashed) normalized to NLO.
\label{fig:phptqs}}
\end{figure}

\begin{figure}[tbh]
\epsfxsize=3in 
\centerline{\epsfbox{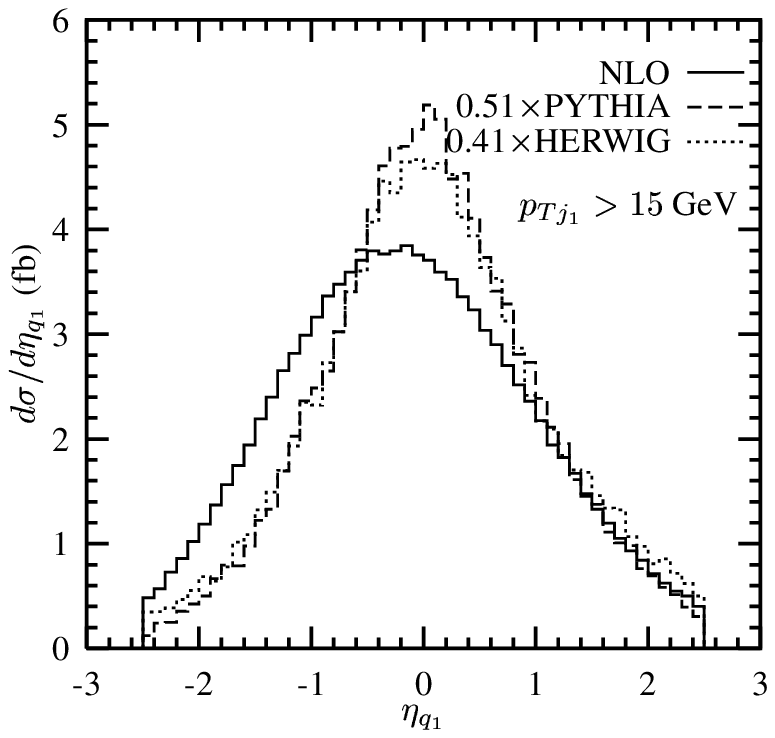}}
\caption{Pseudorapidity $\eta_{q_1}$ of the observed non-$b$ jet in the
$t+1$ jet final state (``jet veto'') of $s$-channel production from NLO
(solid), PYTHIA (dashed), and HERWIG (short-dashed) normalized to NLO.
\label{fig:phetaqs}}
\end{figure}

\begin{figure}[tbh]
\epsfxsize=7.25in 
\centerline{\epsfbox{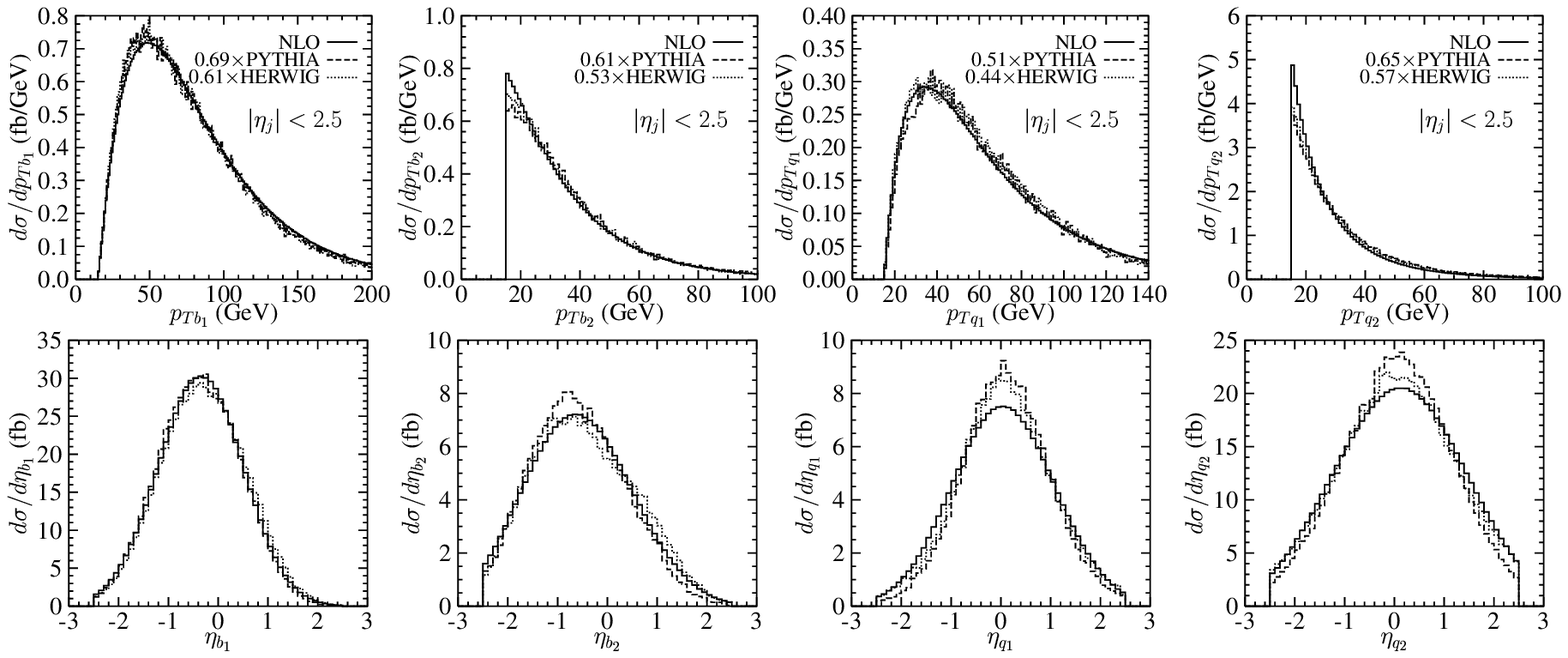}}
\caption{Transverse momentum $p_T$ and pseudorapidity $\eta$ for the
$p_T$-ordered jets $j_i$ in the $t+2$-jet final state of $s$-channel
production from NLO (solid), PYTHIA (dashed), and HERWIG
(short-dashed). $j_1$ and $j_2$ are divided into $b$ jets $b_i$ and
non-$b$ jets $q_i$.  Curves are normalized to NLO in each column.
\label{fig:phjs}}
\end{figure}

The $s$-channel cross section is not well represented by either HERWIG
or PYTHIA.  In Figs.~\ref{fig:phptbs}--\ref{fig:phjs} we see that the
event generators \emph{overestimate} the number of events in all final
states that include additional radiation, and \emph{underestimate} the
number in the $t+b$-jet ($Wb\bar b$) final state.  The origin of this
artifact in PYTHIA is that $s$-channel production is implemented by
producing a $W^{\prime}$ that is set to have the same mass as a real
$W$ boson.  The problem is that the scale is hard-coded to be the
$W^{\prime}$ mass.  Hence, this extra radiation comes from using too
small a scale.  The scale should be something on the order of the
invariant mass of the $t\bar b$ pair \cite{Harris:2002md}, or at least
the top-quark mass.  HERWIG also uses a scale that is too small, but
it uses different scales in the hard cross section for different
events.  For example, gluon-initiated events use an average hard scale
around 100 GeV, and sometimes use a hard scale less than 1 GeV.  This
artificially overweights those events which have additional radiation.

The shapes of the transverse momentum spectra are reasonably well
modeled by both event generators, but additional radiation is too
sharply peaked at small pseudorapidity.  Hence, analyses with tighter
cuts than shown in Figs.~\ref{fig:phptbs}--\ref{fig:phjs} will
demonstrate a larger discrepancy.  The net effect is that $s$-channel
production in the $tb$-jet ($Wbb$) final state has been underestimated
by the event generators by at least a factor of 1.32--1.45; whereas in
the $t+1$ non-$b$-jet final state, the signal is overestimated by at
least a factor of 2--2.4.  In the $t+2$ jet final state, both
generators overestimate the number of events by at least a factor of
1.4--2.3 depending on the exact final state.  Since the discovery
channel is $Wb\bar b$, it should be easier to find $s$-channel
production than previously estimated by the experiments.  However,
this means that the $s$-channel contribution to $Wb\bar b$ as a
background to low-mass Higgs production has been underestimated by the
same amount.

\subsection{Correcting the event generators}
\label{sec:gencor}

The first general-purpose next-to-leading-order showering event
generators for hadron colliders are beginning to appear
\cite{Frixione:2002ik}.  However, they continue to rely on the ability
to cleanly separate initial-state and final-state radiation.
Single-top-quark production through the $s$-channel exhibits such a
separation, but has not yet been added to the programs.  A corrected
leading-order generator using a $K$-factor is expected to perform well
enough.  On the other hand, there is no clear separation in
$t$-channel production between initial-state and final-state
radiation.  Given the similarity to deep-inelastic scattering, it may
be possible to construct a next-to-leading-order event generator for
$t$-channel production \cite{Collins:2000qd}, but no working
implementation exists.  Nevertheless, in order to compute detector
efficiencies it is necessary to have predictions at the level of
detectable particles (pions, photons, etc.).  Therefore, I propose a
simple matching prescription that bypasses many of the problems
\cite{ZStalkA}.

The goal is to produce a set of events with the proper shape and
normalization.  The most important thing to match is the shape of the
distributions in a given final state.  Once the shapes are matched,
generated events may be multiplied by an overall probability that is
normalized by the NLO calculations after cuts.  This prescription is
based on the essential understanding that the next-to-leading-order
calculations are jet calculations.  Therefore, matching must be done
at the jet level.  This is accomplished in four steps.
\begin{enumerate}
\item Generate events with MadEvent \cite{Maltoni:2002qb} or CompHEP
\cite{Boos:2004kh} using the tree-level Feynman diagrams that produce
all measurable final states.  The $t$-channel parton-level processes
for top-quark production are $qb\to tq^\prime$, $qb\to tq^\prime g$,
$gb\to t\bar q q^\prime$, and $qg\to tq^\prime \bar b$, where $q$ and
$q^\prime$ stand for the five light quarks.  The $s$-channel
parton-level processes are $q\bar{q}^\prime\to t\bar b$,
$q\bar{q}^\prime\to t\bar bg$, $qg\to t\bar bq^\prime$, and
$g\bar{q}^\prime\to t\bar b\bar q$.  Use the DDIS scales ($\mu_l=Q^2$,
$\mu_h=Q^2+m_t^2$) for $t$-channel production, and $M_{tb}$ for
$s$-channel production to minimize changes in the shape between LO and
NLO.
\item Feed these events into HERWIG or PYTHIA with full showering turned on.
\item Create four samples ($tj$, $tb$, $tjj$, and $tbj$) for
$t$-channel production with minimal cuts, and a well-defined jet
definition, e.g., $k_T$ clustering with $\Delta R<1$, applied to
stable particles in the event record ($\pi^\pm$, $\gamma$, $e^\pm$,
$\mu^\pm$, etc.).  $s$-channel production will have three samples
($tb$, $tj$, and $tbj$).
\item Produce the same final states with the NLO calculation, and use
the same cuts and jet definition.  Normalize the weights of the events
in each of the event generator samples to the cross sections of these
NLO samples.  Remember that the NLO jets should be $E_T$-ordered, but
that explicit heavy-flavors in the final state must be tracked.
\end{enumerate}

In principle we would want to apply the jet definition to the event
generators at the point of clustering.  However, this is not possible
in PYTHIA because of the order in which events are processed.
Furthermore, both event generators do some slight reshuffling and
energy rebalancing at a later point in the process.  The NLO
calculation is not aware of long-range final-state effects, but
constraints in the event generators on the direction of particles
coming from the clustering stage maintain the correspondence with the
perturbative calculation.  Matching to jets produced out of stable
particles has been found to work to better than 1\% (the limit of the
statistical accuracy checked) for the simpler case of $s$-channel
$W^{\prime}$ production \cite{Sullivan:2002jt,Sullivan:2003xy}, where
the event generators perform well.  In general, a systematic study of
the uncertainty in the matching should be performed for each set of
cuts.

\section{Conclusions}
\label{sec:conclusions}

Precise measurements of single-top-quark production in both the
$s$-channel and $t$-channel modes will begin to appear within the next
few years using data from run II of the Fermilab Tevatron, and
eventually the CERN LHC.  These cross sections are distinguished by
the frequency of $b$ jets occurring in the $W+2$ jet sample, with
$Wb\bar b$ ($Wbj$) most often coming form $s$-($t$-)channel
production.  These measurements will provide direct extraction of the
CKM matrix element $V_{tb}$, and may allow discovery and
interpretation of new physics.

Historically, the ability to interpret the measurement of the
single-top-quark cross sections was expected to be dominated by
theoretical uncertainties.  In Sec.~\ref{sec:uncertainties}, we see
that these uncertainties, while still large, are under much better
control.  The uncertainty in the top-quark mass completely dominates
the prediction for the $s$-channel cross section, but this will
improve with additional data.  The measurement of the $s$-channel
cross section will be statistics limited at the Tevatron, and
difficult to distinguish from backgrounds at the LHC
\cite{Stelzer:1998ni}.  Hence, the theoretical uncertainty in
$s$-channel production should be smaller than the expected
experimental accuracy at either a high-luminosity Tevatron, or the
LHC.

Current theoretical uncertainties in $t$-channel production limit
extraction of $V_{tb}$ to about $\pm 8\%$.  The next-to-leading-order
distributions presented in Secs.~\ref{sec:etvsnlo} and
\ref{sec:gentchan} improve the kinematic uncertainty in $t$-channel
production, so that parton distribution function uncertainties will
dominate our ability to interpret the data.  This uncertainty could be
reduced by measuring the ratio of $t$-channel production to $Zb$
production.  However, the latter cross section is small, and thus we
should consider alternate possibilities, such as $Wc$ or $Wj$
exclusive final states.  Any improvement in the measurement of the $b$
or gluon distributions at scales near the top-quark mass, and proton
momentum fraction $x$ near $0.1$, will directly reduce the remaining
theoretical uncertainty.

The availability of fully differential next-to-leading-order
calculations for hadronic processes has begun a new era of precision
comparisons between theory and experiment.  In order to fully realize
the potential of these theoretical improvements, we must first
understand how to interpret the calculations.  In
Sec.~\ref{sec:nlopred}, we see that NLO calculations produce
\emph{jet} distributions in the final state, rather than \emph{parton}
distributions.  This forces us to describe the results in terms of
$E_T$-ordered jets, rather than NLO distributions of a specific quark.
It further ties the theoretical results to the explicit choice of jet
definition.  We see in Sec.~\ref{sec:jetdef} that the effect of jet
definition on the shapes of distributions is larger than all
theoretical uncertainties combined.

Finally, fixed-order calculations are generally compared to showering
event generators, such as PYTHIA and HERWIG, rather than directly to
data, in order to understand detector efficiencies.  Therefore, we must
understand how to compare next-to-leading-order jets to jets produced
from showering.  A simple matching scheme is presented in
Sec.~\ref{sec:gencor}, that uses the NLO calculation to normalize the
measurable event samples.  Ultimately, we would like to have a
next-to-leading-order event generator that can handle initial-state
and final-state interferences.

Now that the theoretical predictions for single-top-quark production
are under control, a few questions remain regarding the decay of the
top quark.  One of the most interesting features of single-top-quark
production is the possibility of measuring the polarization of the top
quark \cite{Czarnecki:1990pe,Mahlon:1996pn,Stelzer:1998ni}.  The
program used to produce distributions for this paper, ZTOP, does not
decay the top quark.  However the complete analytic NLO spin-density
matrices for production do appear in Ref.~\cite{Harris:2002md}, and
could be combined with the known NLO decays from
Ref.~\cite{Czarnecki:1990pe}.

A spin-dependent program would allow an estimate of the uncertainties
associated with using alternate search strategies, such as cutting on
$H_T$, the scalar sum of the transverse energies of the particles in
the event.  It is clear from the distributions in
Sec.~\ref{sec:eventgen}, that the $H_T$ for single-top-quark production
is larger than predicted at LO in the $W+2$ jet inclusive sample.  The
inclusion of top-quark decay is necessary to determine whether this is
a serious problem or not.  Ultimately, a program that includes decays
will be necessary to determine whether the single-top-quark background
to new physics, such as Higgs production, has been greatly
underestimated, or whether specific cuts accidentally reduce the
discrepancies.  For now, however, we have a reliable theoretical
calculation, and method for matching to event generators, that will
allow for accurate interpretation of single-top-quark production data
when it is discovered.

\begin{acknowledgments}
This work is supported by the U.S.\ Department of Energy, 
High Energy Physics Division, under contract DE-AC02-76CH03000.
\end{acknowledgments}

\end{document}